\documentclass[sigconf]{acmart}

\usepackage{float}
\usepackage{subfigure}
\AtBeginDocument{%
  \providecommand\BibTeX{{%
    \normalfont B\kern-0.5em{\scshape i\kern-0.25em b}\kern-0.8em\TeX}}}
\usepackage{tabularx}
\usepackage{bbding}
\usepackage{graphicx}
\usepackage{geometry}
\usepackage{subfigure}
\usepackage{amsmath}
\usepackage{makecell}
\usepackage{float}
\usepackage{color}
\usepackage{booktabs}
\usepackage{caption}
\usepackage{hyperref}
\usepackage{xcolor}

\copyrightyear{2025}
\acmYear{2025}
\setcopyright{acmlicensed}\acmConference[DIS '25]{Designing Interactive
Systems Conference}{July 5--9, 2025}{Funchal, Portugal}
\acmBooktitle{Designing Interactive Systems Conference (DIS '25), July 5--9,
2025, Funchal, Portugal}
\acmDOI{10.1145/3715336.3735787}
\acmISBN{979-8-4007-1485-6/2025/07}

\newcommand{\remove}[1]{}

\newcommand{\add}[1]{{#1}}

\author{Yuqi Wang}
\authornote{Equal contribution.}

\orcid{0009-0006-8323-9800}
\affiliation{Ontario Institute for Studies in Education\\
\institution{University of Toronto}
\city{Toronto}
\country{Canada}}
\email{yuqiw.wang@mail.utoronto.ca}

\author{Sirui Wang}
\authornotemark[1] 
\orcid{0009-0000-9773-856X}
\affiliation{Tisch School of the Arts\\ 
\institution{New York University}
\city{New York}
\country{United States}}
\email{sw5546@nyu.edu}

\author{Shiman Zhang}
\orcid{0009-0009-3350-8808}
\affiliation{
\institution{University of Edinburgh}
\city{Edinburgh}
\country{United Kingdom}} 
\email{s2658776@ed.ac.uk}

\author{Kexue Fu}
\orcid{0000-0002-2929-2663}
\affiliation{
\institution{Studio for Narrative Spaces\\City University of Hong Kong}
\city{Hong Kong}
\country{China}}
\email{kexuefu2-c@my.cityu.edu.hk}

\author{Michelle Lui}
\orcid{0000-0002-5859-8354}
\affiliation{Ontario Institute for Studies in Education\\
\institution{University of Toronto}
\city{Toronto}
\country{Canada}}
\email{michelle.lui@utoronto.ca}

\author{RAY LC}
\authornote{Correspondences should be addressed to LC@raylc.org.}
\orcid{0000-0001-7310-8790}
\affiliation{
\institution{Studio for Narrative Spaces\\City University of Hong Kong}
\city{Hong Kong}
\country{China}}
\email{LC@raylc.org}

\begin{teaserfigure}
    \centering
    \includegraphics[width= 1\linewidth]{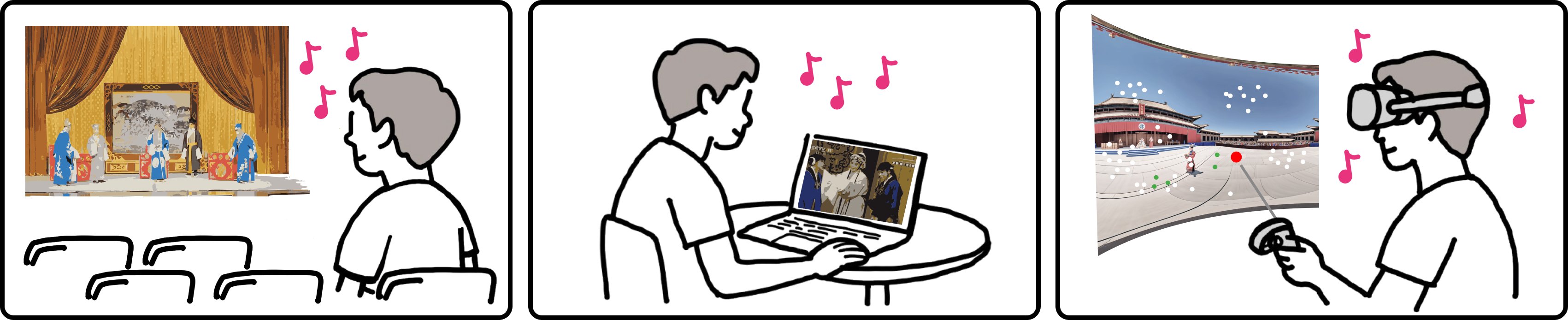}
    \caption{Traditional Performing arts theatre experience versus screen-based video watching experience versus Spatial Interaction-based Segmented-Audio (SISA) prototype experience. \textbf{(Left)} Traditional performance, though sensory engaging and immersive, requires physical attendance at theatre venues, extended passive listening, and relies on conventional practitioner enactment, highlighting fundamental constraints in ICH engagement and preservation. \textbf{(Middle)} Screen-based videos provides temporal flexibility but still lacks immersive engagement essential for authentic cultural appreciation. \textbf{(Right)} Our SISA prototype enables active audience participation through spatial interactions with segmented audio elements in an immersive environment.}
    \label{fig:interaction with SISA system}
\end{teaserfigure}

\begin{document}

\title[From Temporal to Spatial]{From Temporal to Spatial: Designing Spatialized Interactions with Segmented-audios in Immersive Environments for Active Engagement with Performing Arts Intangible Cultural Heritage}

\begin{abstract}

Performance artforms like Peking opera face transmission challenges due to the extensive passive listening required to understand their nuance. To create engaging forms of experiencing auditory Intangible Cultural Heritage (ICH), we designed a spatial interaction-based segmented-audio (SISA) Virtual Reality system that transforms passive ICH experiences into active ones. We undertook: (1) a co-design workshop with seven stakeholders to establish design requirements, (2) prototyping with five participants to validate design elements, and (3) user testing with 16 participants exploring Peking Opera. We designed transformations of temporal music into spatial interactions by cutting sounds into short audio segments, applying t-SNE algorithm to cluster audio segments spatially. Users navigate through these sounds by their similarity in audio property. Analysis revealed two distinct interaction patterns (Progressive and Adaptive), and demonstrated SISA's efficacy in facilitating active auditory ICH engagement. Our work illuminates the design process for enriching traditional performance artform using spatially-tuned forms of listening.

\end{abstract}

\ccsdesc[500]{Applied computing~Arts and humanities}

\keywords{Intangible Cultural Heritage (ICH), Virtual Reality (VR), Spatial Audio, t-SNE, Active Engagement}

\maketitle

\section{Introduction}\label{sec:Introduction}



Much of the richness of our culture comes from sources that cannot be easily documented or quantified, because they are not buildings or objects, but rather cultural practices that define us as a community. Traditional performing arts as a form of Intangible Cultural Heritage (ICH) \cite{unesco2003basictexts}, including music, dance, theatre, pantomime and beyond, is an example. UNESCO recognizes these cultural practices as "invaluable and irreplaceable sources of life and inspiration"  \cite{unesco2003basictexts}. However, ICH performing arts face significant threats due to weakened practice and transmission \cite{Burla2022REVALUATION,Withers1980Unbalanced, Bain1969Performing, unesco_dive_ICHheritage_general_threats}. These threats are intensified by challenges in sustainable generational transmission, as traditional ICH oral instruction methods \cite{unesco2010peking} face increasing difficulty in maintaining continuous practice and knowledge transfer across generations in rapidly modernizing societies. As a consequence, some ICH performing arts events, such as Kun Qu Opera \cite{unesco2008Kunqu} and Peking Opera\cite{unesco2010peking}, that were once frequent and widely attended now occur rarely\cite{unesco_meshrep}.

While museums strive to safeguard cultural heritage, they primarily focus on tangible artifacts, such as historical buildings, artworks, photos and models \cite{hksarg2022pressrelease}, often struggling to capture ICH's "living and ever-changing" nature \cite{unesco2023photoexhibition} and integrate it effectively into their activities \cite{Nwabueze2013}. With ICH performing arts facing a steep decline, the sustainability challenges in preserving their auditory component are particularly significant, as traditional preservation methods require extensive resources and location-dependent engagement. Traditional methods of preserving and presenting audio ICH often lead to passive listening experiences for audiences. These methods typically involve extended periods of listening time of over two hours, demanding full attention of the audience without providing engaging interaction in return. The inherently passive, temporal, and linear nature of these artforms often fails to engage modern audiences and younger generations, who have shorter attention spans and prefer interactive, self-expressive content \cite{amtlab2022genz}. As a result, they often struggle to meaningfully engage with, understand, and contextualize the heritage value of these artforms \cite{CerquettiFerrara2018, Brown2005}. 

Current research have popularized the use of Virtual Reality (VR) to enhance engagement with audio-related ICH artforms \cite{Cai2023The, Ch’ng2020The,Tsita2023A,Zhang2023A,Othman2021Overview,Kim2022Aural,arendttorp_ICH_VRgesture,Kong_VRGAME_Diabolo_ICH,Zhangetal_VR_NVSHU_ICH,rao2024formationcreator, Liu_Hua'er}. However, those approaches still adhere to the temporal and linear nature of auditory ICH, often requiring long periods of passive listening. Studies have shown that compared to passive listening, active listening significantly improves engagement\cite{Kliuchko2019Fractionating, Remijn2010Active}. Some recent studies \cite{erol2022soundoff, cowen2018mapping, cowen2020music} provide an inspiring approach to interactive listening experiences. These studies utilized machine learning algorithms, such as t-distributed Stochastic Neighbor Embedding (t-SNE)\cite{van2008visualizing}, to transform temporal audio experiences into active interactions in VR 3D spaces \cite{erol2022soundoff} and interactive spatial online audio  interfaces\cite{cowen2018mapping,cowen2020music}. By clustering audio segments based on their acoustic features, these approaches attempt to make audio exploration more tangible, visually engaging, and spatially interactive. However, novel approaches to interactive listening experiences remain unexplored in the ICH domain.

Inspired by these technological advances and the challenges in ICH engagement, we propose a new design methodology that aims to transform passive temporal auditory ICH experiences into active ones, through our Spatial Interaction-based Segmented-Audio (SISA) prototype. Our approach segments ICH audio recordings into 5-second segments and employs t-SNE algorithm \cite{van2008visualizing} to cluster these segments spatially by their acoustic features, creating navigable 360-degree VR environments that may enable new forms of auditory ICH engagement. Our research examines both the theoretical considerations of spatial-temporal transformation in interaction design and its practical applications for ICH engagement, leading to two research questions:


\textbf{RQ1:} \textit{How do we design engaging auditory ICH interactions utilizing the spatial properties of immersive experiences?} 

\textbf{RQ2:} \textit{What patterns of interaction emerge when users listen to auditory ICH audio segments that are mapped into spatial experiences?}

\begin{figure*}[h]
    \centering
    \includegraphics[width=1\linewidth]{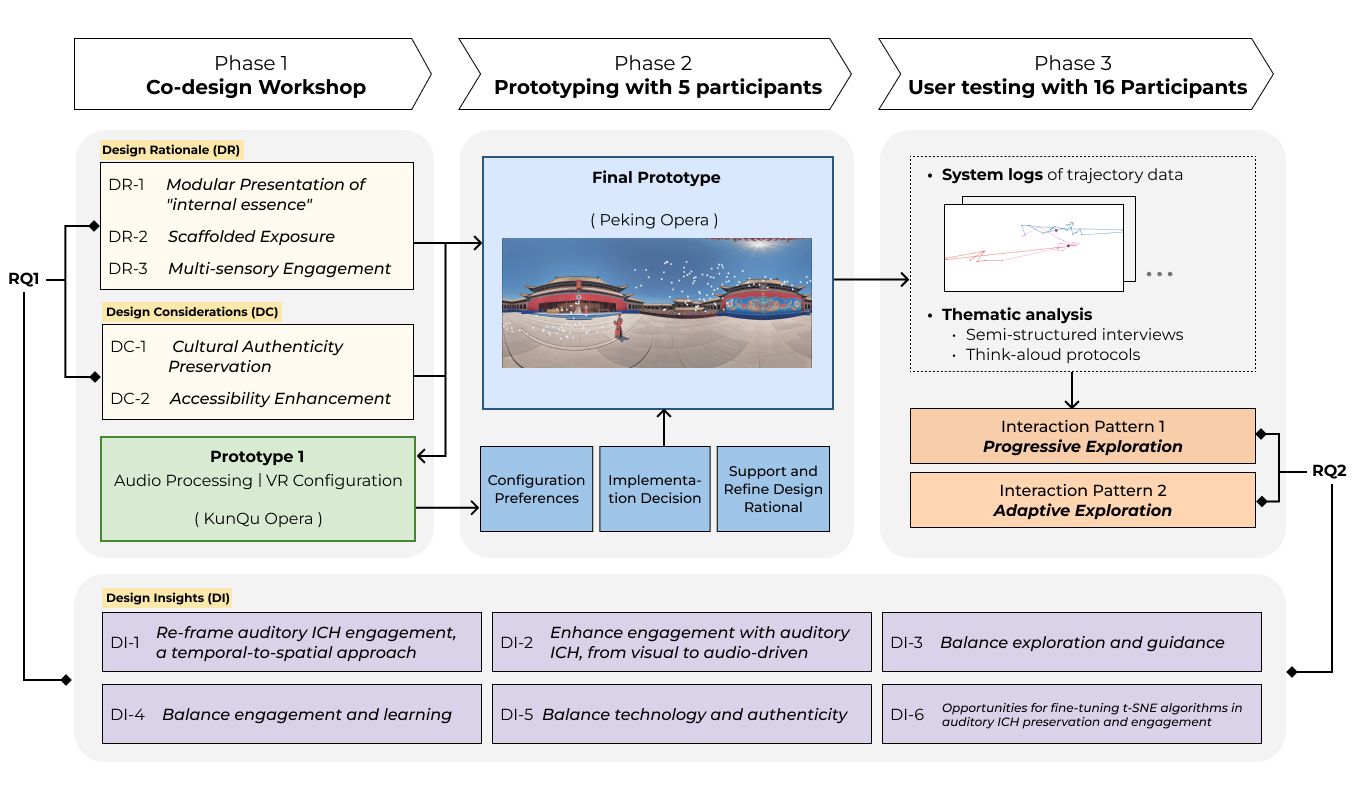}
    \caption{A detailed break down of our project phases.}
    \label{fig:methodoverview}
\end{figure*} 

To address our research questions, we structured our investigation into three sequential phases, each building upon insights from the previous phase to systematically design, develop and refine the SISA system. 
Phase 1 and 2 uncovered critical design considerations for developing spatial transformations of temporal audio (RQ1), while Phases 3 examined how users experience and interact with spatially transformed auditory ICH content in SISA prototypes (RQ2). Refer to Methods section and Figure \ref{fig:methodoverview} for the detailed breakdown of the three phases. Phase 1 employed a participatory co-design approach with seven stakeholders representing diverse perspectives: one designer, one developer, one ICH practitioner, two researchers, and two user representatives. Through collaborative co-design sessions, we identified design rationale and critical design elements for transforming linear audio experiences into interactive spatial arrangements.
Phase 2 consisted of testing prototyping with five participants to validate and refine the design elements identified in Phase 1. This exploratory investigation examined the technical feasibility of Prototype 1 and informed subsequent system refinements. Phase 3 comprised comprehensive user testing with 16 participants to investigate interaction patterns and user experiences of the final prototype. Analysis revealed two distinct interaction patterns: Progressive and Adaptive, demonstrating SISA's efficacy in promoting active ICH engagement.   

While traditional ICH transmission faces constraints of resource intensivity and location dependency, our temporal-to-spatial SISA  methodological framework offers a new accessible, interactive and sustainable way to experience ICH performing arts experiences. The SISA approach establishes a design methodology that reconceptualizes interactive auditory ICH experiences beyond conventional approaches. Our work makes significant societal contributions by contributing both theoretical frameworks for spatial audio interaction design and guidelines for creating sustainable, accessible and engaging auditory ICH experiences to preserve and present "the past"\cite{unesco2003basictexts}. This work illuminates methodologies for enriching traditional performance artforms through spatially-tuned forms of listening.

\section{Background}\label{sec:Background}
\subsection{Intangible Cultural Heritage and Performing Arts}

The performing arts is one of the five domains of ICH. They include both visual and audio components, encompassing traditional theater performances that combine acting, singing, dance, music, dialogue, narration, recitation, puppetry, and pantomime \cite{unesco2003basictexts}. These artforms shape cultural identities and enhance social cohesion \cite{de-Miguel-Molina2021}. Despite their significance, these cultural expressions face numerous modern challenges, such as aging practitioners, diminishing youth interest, hampered transmission, and reduced practice \cite{unesco_dive_ICHheritage_general_threats}. Traditional performing arts particularly illustrate these threats. For example, Kun Qu Opera and Peking Opera's vitality are eroding as they rely heavily on master-student relationships for transmission through oral instruction, observation, and imitation \cite{unesco2010peking}. 
The preservation challenge is further complicated by evolving audience preferences and consumption patterns. Modern audiences, accustomed to interactive and fast-paced digital entertainment, often find traditional performances less engaging \cite{amtlab2022genz}. 

Recognizing the urgency of these challenges, international organizations have developed strategies to safeguard ICH. UNESCO's approach, in particular, advocates for transforming ICH from "fragile" to "truly alive" \cite{unesco2003basictexts}, encouraging the development of global appreciation and active engagement in these cultural practices. UNESCO suggests media, institutions, and cultural industries playing critical roles in developing audiences, raising public awareness, and promoting active participation in cultural expressions \cite{unesco2003basictexts}. However, despite these efforts, auditory ICH remains challenging to engage with actively due to its traditionally passive, linear, and time-bound nature.


\subsection{The Limitations of Passive Reception of Traditional, Linear, Time-bound Auditory ICH}
Auditory ICH traditionally limits listeners to passive reception due to its linear, time-bound nature, constraining active exploration and deep engagement. This passive approach, akin to listening to an unstructured lecture without cognitive engagement, differs significantly from interactive experiences like making music or engaging in dialogue. \add{This distinction between passive and active engagement with auditory ICH can be further understood through Schaeffer's four modes of listening: écouter (listening to identify sources), ouïr (passive perception), entendre (selective hearing), and comprendre (understanding meaning) \cite{Schaeffer_4listeningModes}. Traditional ICH experiences often confine audiences to ouïr mode—passive perception without active interpretation.} The contrast between passive and active listening lies in the level of engagement and the listener's agency to interact with manageable information within their attention span. 
Neuroscientific research underscores the importance of active engagement in auditory processing \cite{Remijn2010Active, Kliuchko2019Fractionating,Qin2023EXPRESS:}. Research has shown that active-response listening significantly enhances cognitive engagement and neural activity compared to passive listening \cite{Remijn2010Active}. 
The advantages of active engagement extend beyond auditory stimuli. A meta-analysis of 128 effect sizes from 33 experiments revealed a small to moderate advantage of active exploration over passive observation in spatial knowledge acquisition \cite{Qin2023EXPRESS:}. This research suggests that hands-on, experiential, interactive approaches to experiencing ICH, rather than passive reception, could be beneficial for understanding and remembering nuances of culture.
These results align with the ICAP (Interactive, Constructive, Active, Passive) framework\cite{Chi2014The}, which posits that learning outcomes improve as engagement progresses from passive to interactive modes. If applied to ICH, this framework could suggest the development of more effective content delivery methods, particularly for modern audiences.
Collectively, these studies emphasize the need to move beyond mere documentation and archiving of auditory ICH. By promoting carefully designed interactive experiences, we can potentially enhance engagement levels and facilitate the transmission of cultural nuances, particularly among modern audiences. Our aim is to develop a new approach that addresses the limitations of traditional, passive methods of engaging with auditory ICH, converting the traditional passive reception of auditory ICH into active interaction experiences.

\subsection{Current Practices of ICH engagement}

We examined current technological practices in ICH engagement and preservation, focusing on audio-related aspects, given the potential of interactive experiences to enhance ICH engagement and preservation. Contemporary ICH engagement and preservation practices increasingly adopts advanced digital technologies, including Virtual Reality (VR) \cite{Cai2023The, Ch’ng2020The, Tsita2023A, Zhang2023A, Othman2021Overview, Kim2022Aural, arendttorp_ICH_VRgesture,Kong_VRGAME_Diabolo_ICH,Zhangetal_VR_NVSHU_ICH,rao2024formationcreator, fu_i_2023}, Augmented Reality (AR) \cite{liu2023digital,Tan_ICH_WebAR,SUN_AR_ICH}, Generative Artificial Intelligence (GenAI) \cite{yao2024shadowmaker, he_i_2025, fu_being_2024, lc_together_2023, lc_time_2024}, interactive physical devices \cite{muntean2017designing}, live streaming technologies \cite{lu2019feel}, and parametric expression techniques \cite{rao2024formationcreator} to enhance public understanding and engagement with ICH by improving accessibility and interactivity. VR has demonstrated significant advantages in ICH interaction and engagement. Its strength lies in creating immersive, spatially interactive experiences that enhance learning and meaning-making in cultural heritage contexts \cite{Cai2023The, Ch’ng2020The, Tsita2023A,Zhang2023A}. 

Audio-related ICH engagement and preservation practices are evolving but still limited. While basic preservation methods \cite{Mulauzi2021Preservation} and digitization efforts \cite{Badrajan2023Safeguarding} are common, researchers are exploring more advanced technologies. These include multimodal VR presentations for aural heritage \cite{Kim2022Aural}, and interactive systems like Virtual Guqin and Mixed Reality Guqin for granting greater accessibility and interest building for traditional musical instruments \cite{Yu_CHI2021_GuQin}. 

In ICH performing arts, various technological approaches are being applied. For dance movements, a VR application designed to enhance the demonstration of dance formations through the FormationCreator system enables real-time modifications via voice commands in a virtual environment \cite{rao2024formationcreator}. VR models are also being used for capturing dance moves and rhythms \cite{Othman2021Overview}. In the realm of songs and music, an interactive Virtual Reality (VR) storytelling project explores the Chinese Hua'er "Baxiguliuliu" song \cite{Liu_Hua'er}. For puppetry, the ShadowStory project uses digital narratives to promote creativity in Chinese shadow puppetry, incorporating both visual and audio elements \cite{lu2011shadowstory}. However, for the auditory components of these technology-enhanced ICH performing arts practices, these approaches still adhere to the temporal, linear, and time-bound nature of auditory ICH and passive reception approach, indicating a need for more interactive and engaging auditory experiences. 

Recent work around interactive listening experiences \cite{erol2022soundoff,cowen2018mapping,cowen2020music} offers an inspiring approach and can be applied for auditory ICH transmission and preservation purposes. These studies have utilized machine learning algorithms, such as t-SNE (t-distributed Stochastic Neighbor Embedding), to transform temporal audio experiences into active interaction experiences. For example, Cowen et al. created interactive online audio interfaces using t-SNE to visualize the complexities of emotions conveyed by brief human vocalizations \cite{cowen2018mapping} and music-associated subjective experiences \cite{cowen2020music}. Similarly, the Sound Of(f) project employed t-SNE to cluster and map sounds into a VR 3D space, allowing audiences to interact freely with audio segments in a virtual reality setting \cite{erol2022soundoff}. These works are inspiring and lead us to investigate the current audio processing practices and techniques to transform temporal audio experiences into active interaction experiences for auditory ICH engagement and preservation.

\subsection{Audio Processing Practices}
Dimensionality reduction techniques are essential in audio signal processing to simplify complex audio data and enable more accessible interpretation and visualization. A number of research \cite{fedden_2017, dupont2013nonlinear, lo2012nonlinear} has focused on mapping high-dimensional audio data onto lower-dimensional spaces to facilitate both visual and statistical evaluation. The typical implementation pipeline involves feature extraction, reshaping, and dimensionality reduction of audio samples. Mel-Frequency Cepstral Coefficients (MFCCs) are widely used for feature extraction, while various dimensionality reduction techniques, including PCA, t-SNE, UMAP, and Isomap, have been evaluated for performance. Among these approaches, the combination of MFCCs for feature extraction and t-distributed Stochastic Neighbor Embedding (t-SNE) for dimensionality reduction has proven particularly effective for analyzing audio signals ~\cite{pal2020comparison}. Studies adopting t-SNE technical approaches for interactive listening experiences \cite{erol2022soundoff, cowen2018mapping, cowen2020music} inspired us to design and develop an interactive system for transforming passive, linear auditory ICH experiences into active and interactive ones.

\section{Methods}\label{sec:Methods}

To systematically investigate the design and development of interactive systems for auditory ICH engagement, we employed an iterative research methodology grounded in participatory design and prototype-based inquiry. Our work resulted in the Spatial Interaction-based Segmented-Audio (SISA) system, which represents the first application of machine learning algorithms to enhance ICH engagement within immersive VR environments.



Drawing from participatory design approaches and iterative prototyping methodologies, we structured our investigation into three interconnected phases to systematically develop and evaluate the SISA system. Each phase built upon insights from the previous one, enabling us to refine both interaction design elements and the underlying technical implementation. We specifically focused on (1) understanding the design considerations for transforming temporal ICH audio into spatial interactions; (2) examining how users navigate and make sense of spatially-structured ICH audio content in the immersive environment; and (3) exploring the broader implications of SISA approach for ICH engagement, transmission and preservation in digital age. These objectives led to two primary research questions:






\textbf{RQ1:} \textit{How do we design engaging auditory ICH interactions utilizing the spatial properties of immersive experiences?} 

\textbf{RQ2:} \textit{What patterns of interaction emerge when users listen to auditory ICH audio segments that are mapped into spatial experiences?}

We initiated the design process with a participatory co-design workshop in Phase 1, bringing together seven stakeholders representing diverse perspectives: one Human-Computer Interaction (HCI) designer, one VR developer, one ICH practitioner, two HCI researchers, and two user representatives. Throughout Phase 1 co-design workshop, we identified critical design rationale and design elements and created Prototype 1 using it as "an experimental component" \cite{wensveenPrototypesPrototypingDesign2014}. This prototype served as a vehicle for inquiry \cite{wensveenPrototypesPrototypingDesign2014}, enabling stakeholders to explore and articulate design considerations for transforming linear audio experiences into interactive spatial arrangements.
Phase 2 study utilized prototype 1 as both a "technology probe" \cite{hutchinsonTechnologyProbesInspiring2003} and a "provotype" \cite{mogensenPROVOTYPINGAPPROACHSYSTEMS1992} to validate and refine the design elements identified in Phase 1 with 5 participants. This exploratory investigation helped verify our design rationale and examine the technical feasibility of implementing spatial audio interaction mechanisms. The findings from this study directly informed system refinements and interaction design decisions for the final prototype.
In Phase 3, we presented the final prototype as a research archetype \cite{wensveenPrototypesPrototypingDesign2014} to conduct comprehensive user testing with 16 participants. Participants explored Peking Opera VR scene while researchers collected data through think-aloud protocols\cite{Charters2003_thinkaload}, semi-structured interviews, and system interaction logs. This phase focused on understanding how users navigate and make sense of spatially arranged ICH audio content in practice. Throughout this process, which served as a vehicle for inquiry \cite{wensveenPrototypesPrototypingDesign2014}, we documented, analyzed, and critically assessed our findings, focusing on research contributions tied to the design and development process rather than just the artifact itself \cite{wensveenPrototypesPrototypingDesign2014}. The resulting design insights are summarized in the discussion section of this paper. Figure \ref{fig:methodoverview} illustrates the systematic progression of our three-phase design methodology, which is elaborated in subsequent sections. This study was approved by and conducted according to the guidelines of the University Institutional Review Board. 


\section{Phase 1 Study: Co-design of SISA system}\label{sec:Phase 1 Study Design of SISA system}
In Phase 1, we initiated our investigation with a participatory co-design workshop, bringing together seven diverse stakeholders to explore the transformation of traditional ICH audio experiences into interactive spatial arrangements. This section details our co-design process, including the participants, workshop formats, and the resultant design framework that informed SISA system development. Through stakeholder engagement in the ideation process, we established the SISA approach that balances technical innovation with cultural preservation while prioritizing user engagement.

\subsection{Participants}
We recruited seven participants representing diverse stakeholder perspectives in the ICH and technology domains through professional networks.  The participants included: one designer with experience in cultural heritage projects, one developer specializing in VR applications, one ICH practitioner with expertise in Peking Opera and Kunqu Opera, two researchers in HCI and ICH preservation, and two user representatives with varying levels of familiarity with traditional Chinese performing arts. This diverse group was intentionally selected to bring multiple perspectives to the design process, particularly balancing technical feasibility with cultural \remove{elements}\add{authenticity.} Details see Table \ref{tab:cod_participant_info}.

\begin{table*}[ht]
\centering
\footnotesize 
\begin{tabular}{c p{5cm} p{5.5cm}}
\hline
\textbf{Co-Design Participant ID} & \textbf{Occupation} & \textbf{Relevance to the Study} \\
\hline
CoD1 & Designer with 1 year experience in cultural heritage projects & Focused on digital preservation of Chinese artifacts and contributed insights on integrating cultural elements into the design \add{specifically regarding visual aesthetics, symbolic movements, and performance conventions unique to Chinese opera traditions}. \\
CoD2 & Developer specializing in VR applications for cultural exhibition contexts & Provided technical feasibility perspectives for immersive technology integration. \\
CoD3 & ICH practitioner with expertise in Peking Opera and Kunqu Opera & Offered detailed knowledge on traditional performing arts for cultural authenticity. \\
CoD4 & Researcher in HCI & Focused on user-centered design principles and interaction patterns. \\
CoD5 & Researcher in HCI and education & Provided insights on instructional design, emphasizing effective content delivery. \\
CoD6 & User representative with familiarity with traditional Chinese performing arts & Contributed as a knowledgeable end-user, highlighting engagement factors. \\
CoD7 & User representative with limited exposure to traditional Chinese performing arts & Provided a perspective on accessibility and initial impressions. \\
\hline
\end{tabular}
\caption{Demographic Information of Phase 1 Co-design Workshop Participants (N=7)}
\label{tab:cod_participant_info}
\end{table*}

\subsection{Workshop Format}
The co-design workshop was structured as a three-hour intensive session employing multiple participatory design techniques to explore the transformation of linear ICH audio experiences into interactive spatial arrangements. The workshop includes a 15-min introduction, a 25-min open discussion, a 60-min design exploration discussion, a 30-min synthesis and refinement session, and a 30-min closing discussion. During introduction, participants were briefed on the project goals and signed consent forms. The context about current challenges in ICH engagement and preservation were provided. Then, an open discussion\add{, employing contextual inquiry\cite{karen2017contextual},} was led by two researchers who ensured balanced participation and comprehensive documentation of emerging themes. This session explored fundamental questions about ICH engagement, cultural preservation, and technological intervention. This semi-structured dialogue allowed stakeholders to share their diverse experiences and perspectives on current ICH engagement methods, establishing a rich foundation for subsequent design activities. Next in the 60-minute design exploration session, participants engaged in collaborative ideation focusing on defining design goal, rationale, considerations, and potential approaches for enhancing auditory ICH engagement. \add{To supplement the contextual inquiry discussions, we conducted case study analysis, where participants examined existing ICH preservation initiatives and interactive systems to identify effective design elements and potential implementation challenges. The case studies included traditional performance documentation methods\cite{unesco2010peking}, existing ICH applications\cite{qin_xianglian_2014,arendttorp_ICH_VRgesture,lu2011shadowstory,yao2024shadowmaker,Yu_CHI2021_GuQin,Liu_Hua'er}, and emerging spatial audio technologies\cite{cowen2018mapping, cowen2020music, erol2022soundoff}, providing a comprehensive contextual foundation for ideation.} Furthermore, the group worked together to consolidate ideas and establish key design elements for the SISA system and Prototype 1. 

\subsection{Data Collection and Analysis}
The entire workshop was audio-recorded and later transcribed. Additional data sources included researchers' field notes documenting key discussion points and emerging themes, photographs of sketches and diagrams created during ideation sessions, and written feedback from participants collected at the workshop's conclusion. Our analysis followed a rigorous qualitative approach based on Braun and Clarke's thematic analysis methodology\cite{BraunClarke2006_thematicanalysis}. 
The emerging themes were developed and refined through two rounds of review. 

\subsection{Phase 1 Study Results: Design Rationale, Considerations and Approach}
Through systematic analysis of co-design workshop data, we developed a comprehensive design framework encompassing high-level design goals, underlying rationale, specific considerations, design approach and their implementation in the SISA system design and Prototype 1.

\subsubsection{Design Goal}
Traditional ICH audio content is typically experienced through passive, linear listening, limiting engagement and potentially hindering cultural transmission. Our core focus is on creating more active and engaging ways for users to experience and listen to ICH audio content. As CoD3 articulated: 
\begin{quote}
\textit{"The goal is to inspire more people to listen to these ICH genres...not just within the Chinese community, who already have a deeper understanding of our art...but also to communicate these artistic values to non-Chinese speakers." (CoD3)} 
\end{quote} 
This insight underscores the need for more accessible and engaging approaches to ICH audio content presentation. Our primary design goal is to enhance auditory ICH engagement by transforming passive linear temporal experiences into interactive spatial arrangements within immersive environments. This transformation aims to (1) enable active exploration of ICH audio content; and (2) foster deeper engagement through spatial-temporal interactions. This approach should create a more engaging pathway for users to listen and explore auditory ICH content while supporting ICH preservation and transmission purposes.


\subsubsection{Design Rationales}
Three fundamental rationales emerged from our co-design discussions and practitioner insights: (1) Modular Presentation; (2) Scaffolded Exposure; (3) Multi-sensory Engagement. 

\textbf{\textit{(DR-1) Modular Presentation of "internal essence".}} ICH practitioner CoD 3 emphasized the modular nature of performing arts, noting that \textit{"each performance genre has its own internal essence, modules and grains...own formula and convention"}. This revealed the inherent modularity in traditional performances. CoD 5 suggested a pedagogical approach through \textit{"breaking the genre into modules and components, and showing what each module means"}. This was echoed by CoD 3, \textit{"when people accumulate this knowledge and listening experience, it becomes easier for them to understand and appreciate the performance genre"}. CoD 3 further suggested enabling SISA system to create what practitioners describe as a \textit{"symbol system that allows better communication"}, and to maintain cultural authenticity while facilitating engagement. Hence, we aim to design and develop SISA system to decompose complex cultural elements into discrete, learnable modules while maintaining their interconnections. This can be achieved by applying t-SNE algorithm to cluster audio segments spatially by their similarity in audio property. This approach should facilitate non-linear exploration of ICH listening experience while preserving the authentic features of the artform.


\textbf{\textit{(DR-2) Scaffolded Exposure.}} Stakeholders suggested considering the cognitive load management of users when them experiencing the cultural contexts and content in the system. Both researchers and practitioners emphasized the need for carefully calibrated complexity levels in cultural content presentation, emphasizing the importance of structured initial exposure and gradual deepening of engagement. As researcher CoD4 emphasized, \textit{"We need to start from the surface level, making the system more accessible, interactive, and fun first"}, while researcher CoD5 advocated for gradual information disclosure: \textit{"Don't overwhelm them with too much information. Let the exploration journey progress bit by bit."}  This perspective was strongly reinforced by designer insights, with CoD1 noting: 
\begin{quote}
\textit{"You can't start by explaining deep theories, everyone would find that boring, especially young people. There are too many things to see in the world today - why would people dedicate time to this?" (CoD1)} 
\end{quote}
Drawing from personal experience, CoD3 illustrated this scaffolded exposure principle: \textit{"I started learning Kun Qu and Peking Opera because I loved the visual elements - the beautiful costumes, the makeup, the headdresses. The music came later."} These converging perspectives from multiple stakeholders underscore the importance of implementing a scaffolded exposure architecture in the SISA system that strategically sequences engagement - beginning with visually appealing and immersive auditory elements before gradually introducing deeper cultural contexts and modules.

\textbf{\textit{(DR-3) Multi-sensory Engagement.}} Stakeholders stressed the need of incorporating various sensory dimensions to create comprehensive cultural experiences. As articulated by ICH practitioner CoD 3: \textit{"Performance arts provide many sensory impacts from hearing, vision, and the whole atmosphere. Audience will build emotional connection within this immersive environment."} User presentative CoD6 further reinforced this approach: 
\begin{quote}
\textit{"Theater art emphasizes immersion and sensory impact. The experience is completely different from watching on a computer screen... Even if you say 'I don't understand what they're singing or what they're saying,' for complete beginners, there's still a lot of sensory information...Music itself carries emotions... lyrics are just one aspect." (CoD6)} 
\end{quote}
Hence, we design our SISA system to combine visual, auditory, and interactive elements to provide the rich sensory and immersive experience of traditional performances in the VR environment.


\subsubsection{Design Considerations}Several key considerations emerged from our co-design process that shaped the SISA system's design: (1) Cultural Authenticity Preservation; (2) Accessibility Enhancement. 


\textit{\textbf{(DC-1) Cultural Authenticity Preservation}}: The ICH practitioner underscored the critical importance of preserving cultural authenticity, describing traditional opera as \textit{"a living fossil of our culture, allowing us to glimpse historical perspectives and cultural evolution (CoD 3)."} At the same time, CoD 3 highlighted the balance between maintaining tradition and fostering innovation, stating, \textit{"There are things that are strictly fixed, and things that are flexible. For instance, there are disciplines and manners you have to follow, but there is room for personal creation and innovation."} Furthermore, CoD3 advocated for innovative approaches to promote engagement, transmission, and preservation within the ICH domain, noting that \textit{"when efforts are made to discover and present the internal essence using different methods, and when it's done with dedication, that's respecting the art form."} This perspective underscores the dual necessity of respecting cultural authenticity while exploring creative and innovative pathways in the design of interactive systems for ICH engagement.

\textit{\textbf{(DC-2) Accessibility Enhancement}}. 
The ICH practitioner CoD 3 identified accessibility as a fundamental challenge in traditional performance contexts, explicitly noting that 
\begin{quote}
\textit{"The main challenge is getting people to come to the theater in the first place. Even for the overseas Chinese community, opera feels distant - they know it's national heritage and it's held in high regard, but they don't understand its internal essence (CoD3)."} 
\end{quote}
This challenge highlights a critical barrier to cultural transmission and engagement in conventional performance settings. The practitioner demonstrated strong support for innovative technological approaches to address these accessibility constraints, particularly emphasizing the potential of digital preservation and virtual engagement platforms. The practitioner also emphasized that traditional performance settings, while valuable, can present significant barriers to new audiences. CoD3 highlighted this challenge: \textit{"Another difficulty is how to help people who don't speak Chinese understand the deeper value of our art, including its cultural significance and abstract elements."} This insight reinforces the need for innovative approaches to cultural transmission that can overcome linguistic and cultural barriers.

\subsubsection{Design Approach}
Drawing upon our established design goals, rationales, and considerations, we developed a systematic approach leveraging MFCC-TSNE algorithms to analyze and cluster segments of traditional performance pieces. To transform temporal performance arts music into spatially interactive listening experiences, we developed a novel method to process traditional performance arts music recordings. This approach segments lengthy audio recordings into shorter, manageable units and distributes them across a 360-degree spatial environment. Using MFCC (Mel-frequency Cepstral Coefficients) combined with t-SNE (t-Distributed Stochastic Neighbor Embedding) dimensionality reduction, we created an explorable soundscape that preserves the essence of the original performances (DR-1 and DC-1) while enabling interactive engagement with both cultural visual and auditory elements in the immersive VR environment (DR-3). 

\add{This approach aligns with Schaeffer's four modes of listening \cite{Schaeffer_4listeningModes}- Écouter (listening to identify sources), Ouïr (passive perception), Entendre (selective hearing), and Comprendre (comprehending meaning)- facilitating users' progression through these hierarchical stages of listening. The scaffolded exposure principle (DR-2) enables users to advance from initial source identification (Écouter) toward more sophisticated comprehension (Comprendre), while multi-sensory engagement (DR-3) creates optimal conditions for developing selective hearing (Entendre) capabilities.}
For the remainder of this paper, we refer to this method as the Spatial Interaction-based Segmented-Audio (SISA) approach. By mapping acoustic features into a 3D virtual space using t-SNE dimensionality reduction, we create a taxonomy of soundscapes that aims to facilitate immersive exploration. \remove{This spatial organization enables users to discover relationships between audio segments and understand the genre's distinctive characteristics through active exploration, rather than passive linear listening.} \add{This spatial organization transforms complex temporal performances into navigable spatial audio segments. It aims to enable users, especially novices, to actively and gradually discover relationships between audio segments and then identify genre-specific characteristics such as vocal techniques, instrumentation patterns, and character roles (DR-2). By shifting from passive linear listening to embodied spatial exploration, the system aims to lower cognitive entry barriers and suggests possibilities for enhancing users’ affective and analytical engagement with the genre — directly addressing challenges\cite{unesco2003basictexts} reported by practitioners in cultural transmission and accessibility (DC-2).} 

\subsection{SISA Prototype 1}
Based on previous design rationale and considerations, the SISA approach emphasizes \textit{modular presentation of "internal essence" (DR-1)}, \textit{scaffolded exposure (DR-2)}, \textit{multi-sensory engagement (DR-3)}, \textit{cultural authenticity preservation (DC-1)}, and \textit{accessibility enhancement (DC-2)} in its design. To implement this approach, the SISA system is designed with two primary components:
\begin{enumerate}
    \item Audio Processing: Implements culturally-authentic audio segmentation and modular organization, enabling scaffolded content exploration while preserving internal essence through audio processing methodology. 
    \item VR Configuration: Delivers multi-sensory stimulation via scaffolded progression mechanisms, integrating cultural authenticity and enhanced accessibility through an immersive environment. 
\end{enumerate}


For our initial prototype development, we selected Kunqu Opera \cite{unesco2008Kunqu} as the representative ICH performing arts genre. Following consultation with Kunqu Opera ICH practitioner CoD3, we identified a specific temporal audio segment from the performance titled "The Peach Blossom Fan, 1699"\cite{kunqu_1699peachblossom}, specifically utilizing the sequence from 0:30 to 0:42 minutes. The Peach Blossom Fan\cite{kunqu_1699peachblossom} by Kong Shangren is a historical play about the doomed love between scholar Hou Fangyu and courtesan Li Xiangjun during the Ming dynasty's fall. Their romance, symbolized by a bloodstained fan, is shattered by political turmoil, reflecting the era's personal and national tragedies. Details of Prototype 1 are provided in the next section.

\subsubsection{SISA Prototype 1: Audio Processing}


\begin{figure*}[h]
    \centering
    \includegraphics[width=0.9\linewidth]{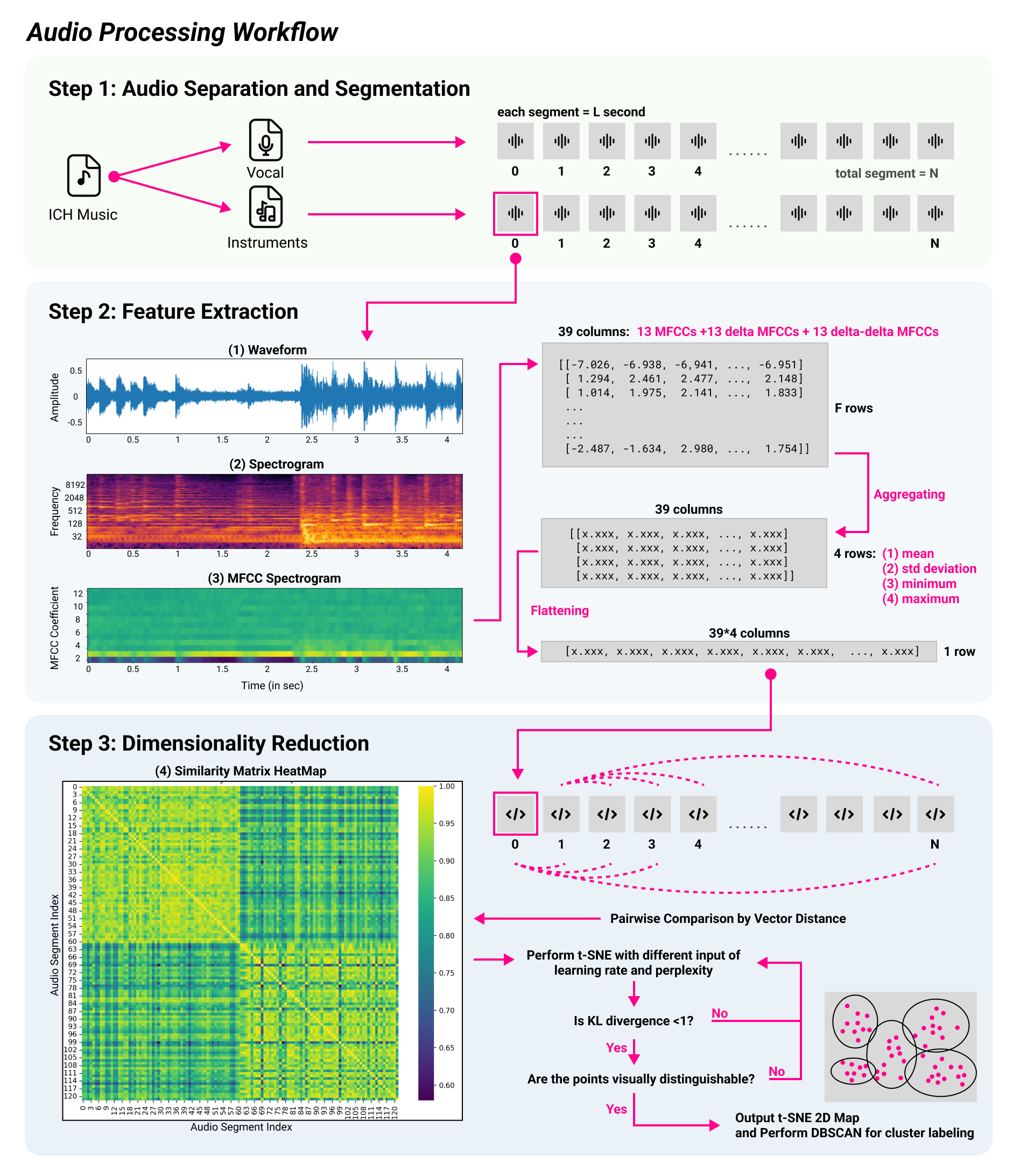}
    \caption{\textbf{SISA Audio Processing Workflow} consists of the following steps: \textit{(1) Audio Separation and Segmentation}, \textit{(2) Feature Extraction}, \textit{(3) Dimensionality Reduction}. In this illustration, the waveform, spectrograms, and the heat map are derived from a short testing clip of Kunqu Opera. }
    \label{fig:audioprocessing}
\end{figure*} 

\paragraph{Audio Separation and Segmentation}
To achieve \textit{modular presentation of internal essence (DR-1)} while preserving \textit{cultural authenticity (DC-1)}, our audio processing workflow begins with careful separation and segmentation of the source material. To prepare the audio data for extraction of MFCC characteristics, we performed a process that involved isolating the different components and segmenting the audio. We first used an open source online AI tool \cite{jameson_2021} specifically designed for audio separation. This tool utilizes machine learning techniques to identify and isolate vocals from audio tracks. By separating these components, we can independently analyze the vocal and instrumental content which can lead to more accurate results when applying t-SNE. With the separated tracks, we used a command-line tool, audio-splitter \cite{Vasilev_2023} to divide each audio file into equally-sized segments. This tool allows for precise control over the length of each segment, which maintains consistency across the dataset. By specifying the desired chunk length, audio-splitter processes the entire audio file and outputs a folder of segments of uniform duration. This segmentation is crucial to ensure that each audio segment is of manageable size for further processing and analysis. Research has shown that a 5-second window size achieves the highest accuracy for pattern recognition using MFCC \cite{4813723}. \remove{Based on these findings, we opted to divide the audio files into 5-second segments when designing prototype 1.} \add{With this technical consideration, we opted to proceed with this 5-second segment length configuration for our Prototype 1.}

\paragraph{Feature Extraction}
In support of \textit{scaffolded exposure (DR-2)} and \textit{accessibility enhancement (DC-2)}, we implemented a comprehensive feature extraction process. MFCC compresses the entire spectrum into a smaller set of coefficients that together represent the overall shape of the spectrum. It converts raw audio signals into computational data. To balance feature richness and performance, the first 13 MFCCs are typically used, as they align with human auditory sensitivity to different frequencies. These are supplemented by Delta MFCCs, which capture changes in cepstral features over time, and delta-delta MFCCs, or acceleration coefficients, which add a longer temporal context \cite{singh2016speaker}. Together, these form a 39-element vector that is assumed to retain sufficient features about the audio sample.

A positive cepstral coefficient suggests that most of the spectral energy is concentrated in the low-frequency regions, while a negative coefficient indicates that the energy is primarily in the high-frequency regions. \add{Consequently, vocalization in elevated frequency ranges characteristic of Kunqu opera, such as the melismatic singing techniques utilized in the "water sleeves" passages of dan role performances, exhibits comparable MFCC patterns that contrast markedly with those in lower frequency ranges or the accompanying ensemble of dizi, pipa, and other traditional instruments \cite{unesco2008Kunqu, dong2014long}. Thus, MFCCs are powerful in facilitating efficacious classification predicated on these spectral attributes}. A corresponding relationship between MFCC coefficients and acoustic features can be observed in Figure \ref{fig:audioprocessing}. To convert audio signals into these coefficients, we used Librosa ~\cite{mcfee2015librosa}, a Python package for music and audio analysis. Librosa generates a coefficient for each frame, resulting in a \((39, F)\) MFCC feature vector for each audio segment, where \(F\) equals number of frames per segment. If we split the original audio sample into \(N\) segments, we will obtain \(N\) number of MFCC feature vectors of size \((39, F)\). 

Since t-SNE takes a 2D array as input, we have to reshape the feature vectors through two steps: aggregation and flattening \cite{pal2020comparison}. We calculated the mean, standard deviation, minimum, and maximum of each feature to aggregate features over frames. This resulted in each feature vector having a size of \((39,4)\). We then flattened all feature vectors into one vector with a size of \((N, 39*4)\), allowing t-SNE to perform dimensionality reduction on all the audio segments. 

\paragraph{Dimensionality Reduction}
To facilitate \textit{multi-sensory engagement (DR-3)} while maintaining \textit{cultural authenticity (DC-1)}, we employed t-SNE dimensionality reduction. This technique allows us to create meaningful spatial relationships between audio segments that reflect their inherent similarities, supporting intuitive exploration of the cultural material. t-SNE is applied to this \((N, 39*4)\) vector where each row corresponds to a high-dimensional data point, and each column represents a feature. This process will result in a two-dimensional coordinate map where similar-sounding segments are positioned closer together \cite{van2008visualizing}. In our case, t-SNE processes the \((N, 39*4)\) audio feature vector by calculating pairwise similarities between audio segments in high-dimensional space, creating a similar distribution in low-dimensional space, and iteratively minimizing the Kullback-Leibler (KL) divergence between these distributions \cite{kullback1951information}. We use the scikit-learn's tool \cite{abraham2014machine} to perform t-SNE. The results vary between runs, so we tested various songs and genres to explore how to optimize results. We observed that (1) a minimum similarity across segments < 0.9 and (2) KL divergence < 1, ideally < 0.5, can meet our expectation. We therefore tried with different combinations of parameters (\(perplexity\) and \(learning\_rate\)), compared multiple runs, and visually evaluated the results to determine which one to adopt. Please refer to Figure. \ref{fig:audioprocessing}  for a detailed visual representation of the complete audio processing workflow.

To better understand the audio features in our t-SNE result \add{and ensure \textit{cultural authenticity (DC-1)} is preserved throughout audio processing}, we \remove{used} \add{implemented} a density-based clustering algorithm. We selected Density-Based Spatial Clustering of Applications with Noise (DBSCAN) among various clustering algorithms because it effectively identifies clusters of arbitrary shapes and it is powerful enough to handle small datasets like ours \cite{kanagala_jaya_rama_krishnaiah_2016}. Given that DBSCAN requires the specification of key parameters—the minimum number of points per cluster (\(min\_samples\)) and the maximum distance between two points (\(eps\)). We fixed \(min\_samples\) to the default value of 5 and ran the algorithm across a range of \(eps\) values to maximize the number of distinct clusters while minimizing noise points. \remove{Then, each resulting cluster was manually reviewed to identify its characteristics and annotated with its feature category based on the UNESCO description. Noise points identified by DBSCAN were also manually labeled with the most appropriate cluster. Refer to Figure.\ref{fig:tsnemapping} for the cluster labeling of selected Kunqu Opera music used in prototype 1.} \add
{Then, we meticulously examined the clusters generated by DBSCAN to characterize their distinct features and labeled according to UNESCO descriptions \cite{unesco2008Kunqu}, with additional validation from CoD3. Our annotations revealed that the clusters naturally aligned with traditional Kunqu opera character roles: young male lead, female lead, and elderly male, which are quintessential to this art form. Notably, we also identified a cluster containing purely instrumental passages without vocals, another defining characteristic of Kunqu opera. Any noise points detected by DBSCAN were manually assigned to their most suitable clusters. Figure \ref{fig:tsnemapping} illustrates the cluster categorization of the selected Kunqu Opera musical segments utilized in prototype 1.}

This clustering approach was instrumental in preparing for our user study analysis. By mapping the spatial distribution of the audio features, we established a structured framework for analyzing user interactions with ICH content in our SISA prototype. The labeled cluster maps would later serve as a reference for overlaying user trajectories, enabling us to discern patterns in audio landscape navigation and reveal insights into exploration strategies and preferences.

\subsubsection{SISA Prototype 1: VR Configuration}
\paragraph{Spatial Mapping}
\begin{figure*}[h]
    \centering
    \includegraphics[width=1\linewidth]{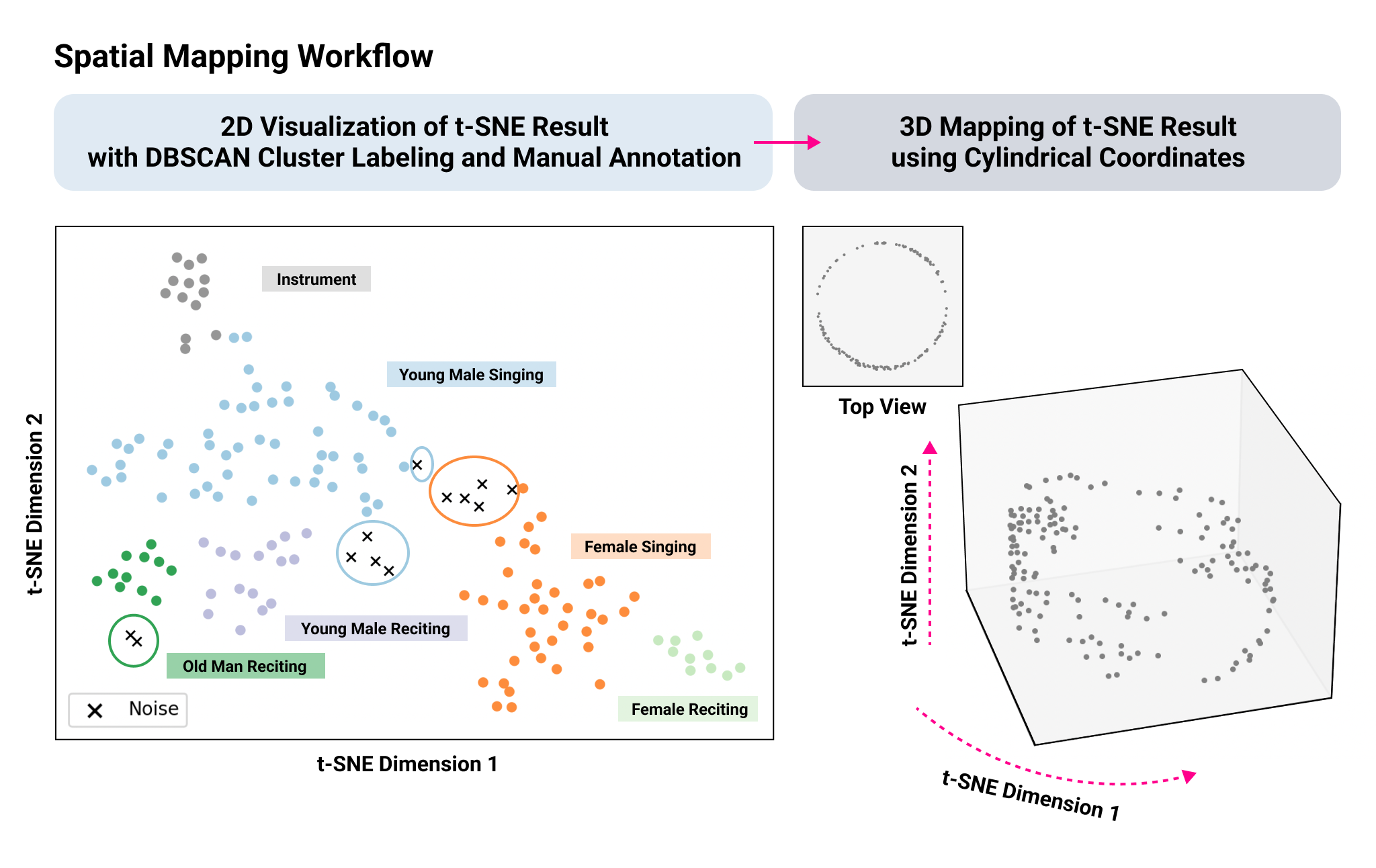}
    \caption{\textbf{SISA Spatial Mapping Workflow. (Left)} 2D Visualization of t-SNE Result with DBSCAN Labeling and Manual Annotation. For the selected Kunqu Opera music, t-SNE effectively grouped similar features together. Using DBSCAN, six distinct clusters were identified, differentiating between instruments and character roles, and whether the performance involved singing or reciting. \textbf{(Right) }3D Mapping of t-SNE Result and its top view. Using a cylindrical coordinate system, the 2D t-SNE results were transformed into a 3D space, ensuring that when the user is positioned at the center, they are equidistant from all points.}
    \label{fig:tsnemapping}
\end{figure*} 
To implement \textit{multi-sensory engagement (DR-3)} and support \textit{scaffolded exposure (DR-2)}, we translated the t-SNE result into navigable 3D space. We extend the 2D t-SNE coordinates (\(x_{2D}, y_{2D}\)) into 3D space by introducing a z-coordinate. This 3D representation maintains the original t-SNE relationships. We set \(z_{3D}\) equal to \(y_{2D}\) and use a fixed radial distance \(r\) to calculate new \(x_{3D}\) and \(y_{3D}\) coordinates as follows \cite{georgious2022scalar}.
\begin{center} 
\(\theta= 2\pi*\frac{(x_{2D} - x_{min})}{(x_{max}-x_{min})}\)\\
\(x_{3D}= r * cos(\theta)\), \(y_{3D}=r* sin(\theta)\), \(z_{3D}=y_{2D}\)\\
\end{center}

The spatial mapping workflow can be visualized as shown in Figure \ref{fig:tsnemapping}.

\paragraph{Visual Environment}
\begin{figure*}[h]
    \centering
    \includegraphics[width=1\linewidth]{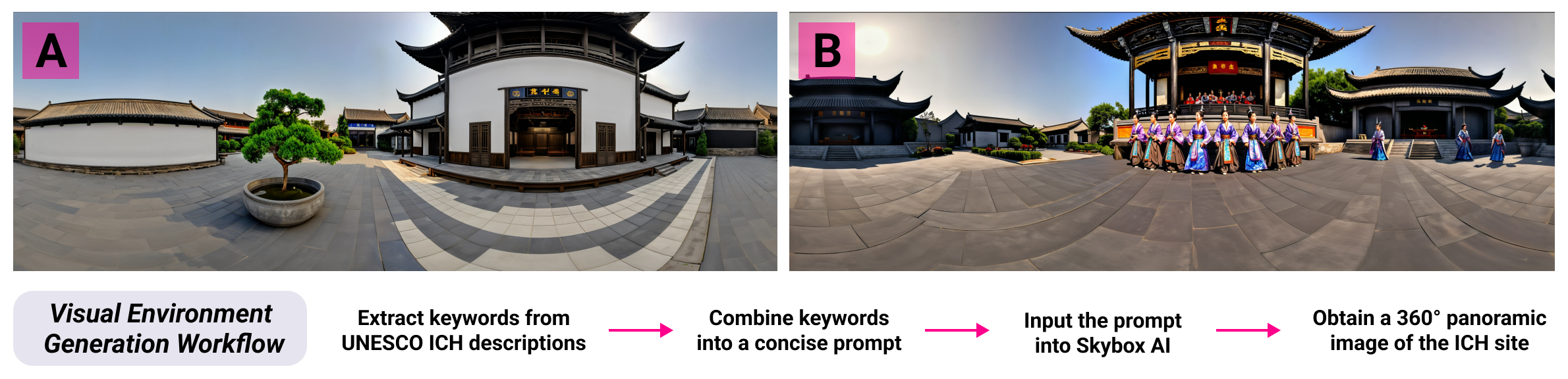}
    \caption{\textbf{SISA Visual Environment Generation Workflow}. We used Skybox AI to generate a visual environment of the ICH scene. The process begins with extracting keywords from UNESCO genre descriptions, followed by combining these keywords into a Skybox-specific prompt. Skybox AI then uses this prompt to create the 360° panoramic image, resulting in a detailed visual representation of the ICH site. 
    \textbf{(A)} Panoramic image of Kun Qu Opera with the absence of people. \textbf{(B)} Panoramic image of Kun Qu Opera with the presence of people.}
    \label{fig:skyboxvisualflowchart}
\end{figure*} 

To enhance \textit{cultural authenticity (DC-1)}, and support \textit{multi-sensory engagement (DR-3)} and visual immersion, we use a generative AI tool, Skybox AI \cite{skybox2024}, to create a culturally and contextually relevant 360-degree image for the VR environment. We used ChatGPT to extract key terms from UNESCO descriptions of the selected ICH genre sites and refine them into a concise prompt. This refined prompt was then input into Skybox AI to generate a panoramic image representing the topic associated with the selected genre. \add{The use of panoramic visual environments aims to achieve contextual fidelity and cultural authenticity (DC-1), addressing the critical role that spatial surroundings play in framing audience interpretation of audio segments.This visual-auditory integration supports users in constructing meaningful relationships between modular audio elements (DR-1) and the holistic cultural experience, facilitating comprehension of the genre's distinctive characteristics through multi-sensory channels (DR-3).} Refer to Figure.\ref{fig:skyboxvisualflowchart} for the visual generation workflow.

\paragraph{Interaction Design} Users can interact with the VR environment with rotational movement, employing three degrees of freedom \cite{googlevr_2019} (rotation, yaw, and roll). As users direct their controller toward these points, they trigger the playback of the corresponding audio segments. To enhance participants' understanding of their exploration progress, when a participant interacts with an audio segment represented by a specific shape and the color of the shape changes from white to red while the segment is playing. Once the segment finishes, the color changes to green, providing visual feedback to indicate that the point has been activated and explored.

\paragraph{Implementation}
The technical implementation focuses on \textit{accessibility enhancement (DC-2)} through an intuitive VR interface. A-Frame \cite{a-frame_2021}, a web-based VR framework, was used to create the SISA environment by placing 3D coordinates within a generated visual environment. Audio segments were assigned to spherical points and played from their positions when selected, utilizing A-Frame's built-in spatial audio functionality.

\section{Phase 2 Study: Support of Design Rationale}\label{sec:Phase 2 Study: Support of Design Rationale}

To validate our design approach and gather insights for future iterations, we recruited five participants to use Prototype 1 to understand user preferences on certain configuration settings of SISA system and collect feedback to support our design rationale. The study focused on investigating two primary aspects: Audio Processing Preferences and VR Configuration Preferences.

\paragraph{Audio Configuration Preferences.}
We used t-SNE to reduce the properties of audio from our selected ICH performing arts genre - Kunqu Opera- creating clusters based on these properties. To better understand the feasibility of presenting these clusters in the ICH context regarding the algorithm's specific configurations, we aimed to understand participants' system preferences regarding:

\begin{enumerate}
 \item \textit{Mixed vs. Split Audio:} Whether participants preferred audio that was not separated into different tracks ("mixed") or audio that was separated ("split").
  \item \textit{Audio Segment Duration:} Whether the current 5-second audio segment duration was sufficient for content consumption and exploration.
\end{enumerate}

\paragraph{VR Configuration Preferences.}
\begin{enumerate}
 \item \textit{Visual environment:} We aimed to collect participants' preferences regarding the absence (A) or presence (B) of people in these images. We generated two 360-degree images based on our prompts using SkyboxAI. Please refer to Figure. \ref{fig:skyboxvisualflowchart} for details.
 \item \textit{Interaction:} Whether participants preferred audio points that changed color after interaction or those that remained unchanged.
\end{enumerate}

\subsection{Participants and Procedure}
We recruited five participants to explore our Prototype 1 featuring Kunqu Opera songs in 5-second audio segments, as suggested by Beritelli and Grasso\cite{4813723}. Participants were asked to think-aloud\cite{Charters2003_thinkaload} during navigation, and we conducted open-ended interviews afterwards for in-depth insights. The thematic analysis of the interview transcriptions are conducted by two researchers. The themes were revised and discussed with the whole research team until final consensus was achieved\cite{BraunClarke2006_thematicanalysis}.

\subsection{Data Analysis and Results}
Thematic analysis of interview transcriptions showed that participants showed a strong preference (four out of five) for the mixed audio configuration, describing it as richer and more immersive. While the distinction between close and far sound was less significant, they appreciated the ease of interaction with larger, closer audio points. Additionally, the interaction mode, where audio points changed color after interaction, was overwhelmingly favored (five out of five) as it provided clear feedback on their exploration progress. This finding strongly supports our design rationale for \textit{multi-sensory engagement (DR-3)} through integrated auditory experiences, while the color-changing feedback mechanism validates our approach to \textit{scaffolded exposure (DR-2)}.  However, participants (three out of five) expressed that the 5-second audio segments were too short, as they found it difficult to grasp the content within such a brief duration. They suggested extending the duration of audio segments, such as to 10-second, for better content consumption and more thorough exploration. This insight particularly informs our \textit{modular presentation approach (DR-1)}, suggesting the need for appropriate temporal granularity in content modules to ensure effective cultural transmission. 

 Thematic analysis of interview transcriptions showed that most participants (four out of five) preferred images that included people, as they felt this added a sense of culture and authenticity to the environment. This preference aligns with our design constraint of \textit{cultural authenticity preservation (DC-1)}, emphasizing the importance of human elements in cultural representation. However, once they realized the images were AI-generated, concerns about authenticity emerged, with some participants expressing doubts about the accuracy of the representations. This finding highlights the tension between technological innovation and cultural authenticity, requiring careful consideration in our implementation of \textit{accessibility enhancement (DC-2)}.

These empirical findings directly supported and refined our initial design rationales while informing specific implementation decisions and configuration preferences for subsequent prototype development. The study results led to the implementation of mixed audio as the default configuration, emphasizing immersive experience and cultural authenticity. Additionally, we integrated interactive feedback to enhance user engagement and exploration awareness. In selecting AI-generated visual elements, we took cultural authenticity into account as part of the design process.

\section{Phase 3 Study: User Interaction Patterns}\label{sec:Phase 3 Study: User Interaction Patterns}
\begin{figure*}[h]
    \centering
    \includegraphics[width=1\linewidth]{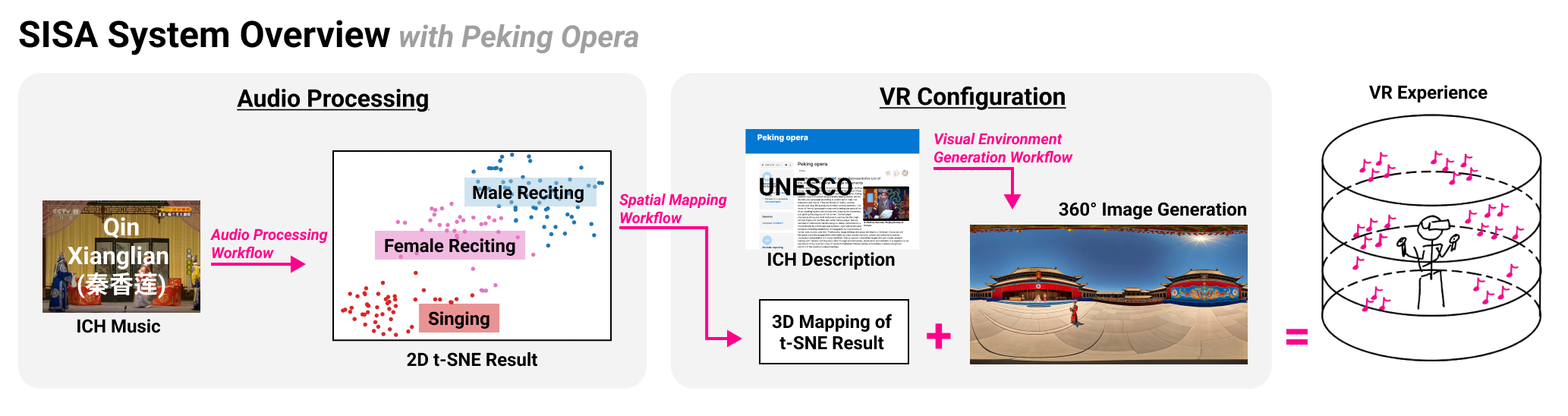}
    \caption{\textbf{System Overview of the SISA} illustrated with Peking Opera genre. The \textbf{audio processing} takes the ICH music as input and perform t-SNE algorithms after segmentation and MFCC feature extraction. The 2D result is then translated into \textbf{3D mapping} and with the generated \textbf{virtual environment} for a culturally relevant immersive environment.  Both components feed into the SISA system giving us a VR experience that enable to spatially explore the ICH performance arts.}
    \label{fig:overview}
\end{figure*}

Building on insights from phase 1 and 2 studies, we refined our prototype design for final prototype and conducted comprehensive user testing with 16 participants to investigate spatial interaction patterns with this prototype presenting Peking Opera\cite{unesco2010peking}. 
This phase 3 study aimed to uncover how users navigate and make sense of spatially transformed ICH content. Please refer to Figure \ref{fig:overview} for the final prototype design workflow chart. 

\subsection{SISA Final Prototype}

For the genre analysis, we selected Peking Opera. Reasons are presented as follows. First, this genre represents distinct linguistic traditions within China: Peking Opera is performed in Mandarin, China's official language. Second, it exhibits notably different levels of exposure among the general Chinese population. Peking Opera, while not actively sought out by younger generations, maintains broader visibility through national media platforms such as the Spring Festival Gala. Third, Peking Opera was inscribed on the Representative List of the ICH in 2010 \cite{unesco2010peking}. Finally, this genre originates from China, making it particularly relevant for our user study focused on participants with Chinese cultural backgrounds.

Peking Opera represents a highly stylized form of traditional Chinese theater that combines music, singing, acrobatics, and elaborate costumes. Its distinctive performance style features unique vocal techniques and orchestral compositions, utilizing traditional instruments such as the thin high-pitched jinghu, the flute dizi, percussion instruments like the bangu and daluo. The genre is renowned for its sophisticated integration of various artistic elements and its rich historical narratives. 

For the audio processing of Peking Opera, we selected the video titled "Qin Xianglian" \cite{qin_xianglian_2014} which tells the tragic story of Qin Xianglian seeking justice after being betrayed by her husband. We used SISA audio processing workflow (Figure \ref{fig:audioprocessing}) to work on the selected audio clip and split it into 10-second segments. \add{The extension from 5-second segments in Prototype 1 to 10-second segments in the SISA Final Prototype reflects an optimization decision aimed at balancing fine-grained exploration with meaningful comprehension. This adjustment preserves the internal essence of the original performance components (DR-1 and DC-1) while supporting cognitive manageability (DC-2)—enabling users to identify character types, instrumental patterns, and emotional qualities that define Peking Opera through appropriately sized modular audio segments (DR-1 and DR-2). For instance, considering the line \cite{qin_xianglian_2014}:
\begin{quote}
Turning my head, I will say to Xianglian: (timestamp 00:24:55-00:25:00)

I advise you to shatter your empty hopes (timestamp 00:25:00-00:25:05)
\end{quote}
In our previous 5-second configuration, this meaningful lyric would be fragmented, potentially confusing users encountering only partial phrases. The 10-second segment would provide more sufficient contextual integrity for proper comprehension while still maintaining manageable cognitive processing. Based on feedback from our Phase 2 Study highlighting preference with mixed audio and the need for longer segments, we refined our parameters when implementing, we modified the configuration in applying SISA Audio Processing and} \remove{We applied SISA} Spatial Mapping Workflow (Figure \add{\ref{fig:audioprocessing} \&} \ref{fig:tsnemapping}) to map the audio segments in the virtual environment. We generated visual environment using SISA Visual Environment Generation Workflow (Figure \ref{fig:skyboxvisualflowchart}) \add{, incorporating the preference of having human figures within the scene}. The system overview of SISA Final Prototype is illustrated in Figure \ref{fig:overview}. \add{The clustering results for Peking Opera in our final prototype, similar to our Prototype 1, revealed characteristic patterns consistent with UNESCO documentation \cite{unesco2010peking}. Unlike Kunqu Opera, Peking Opera demonstrated distinctive acoustic distribution patterns, particularly in how instrumental sections integrated with character performances within the cluster space when using the mixed audio configuration. This comparison illuminates how different operatic traditions establish unique structural relationships between vocal and instrumental elements, suggesting possibilities for genre-specific spatial mapping approaches in future applications, as elaborated in our discussion section.}

\subsection{Participants}
We conducted the user study of our final prototype with 16 participants with divergent knowledge levels of Peking Opera, age groups (11 participants aged 20-27, 4 participants aged 28-35, and 1 participants aged 35+), gender (4 female and 12 male), cultural background (16 with Chinese background, which 11 with mixed cultural backgrounds, 5 with solely Chinese background), and mixed VR usage history. Table \ref{tab:participant_info} provides an overview of the participants' demographics, along with their prior experience with Peking Opera. Each participant was equipped with a VR headset (Quest 2) and an iPad to complete surveys, a phone to record videos, and another phone to record think-aloud\cite{Charters2003_thinkaload} and interview audio. Prior to participation, each participant provided informed consent, acknowledging that their survey responses, interview data, and videos during the workshop would be collected anonymously for analysis and that they could withdraw from the study at any time for any reason.

\begin{table*}[ht]
\centering
\footnotesize 
\begin{tabular}{c c c p{3cm} p{2.9cm} p{4.4cm}}
\hline
\textbf{Participant} & \textbf{Gender} & \textbf{Age Group} & \textbf{Cultural Background} & \textbf{Experience with Peking Opera} \\
\hline
1 & Male & 20-27 & Chinese, American & Once or twice in a lifetime. \\

2 & Female & 28-35 & Chinese, Canadian, American & Never heard of    Peking Opera. \\

3 & Male & 28-35 & Chinese, Canadian, American &   Never heard of    Peking Opera. \\

4 & Female & 20-27 & Chinese, American &   Frequently. More than 10 times in a lifetime. \\

5 & Male & 20-27 & Chinese &   Never heard of    Peking Opera. \\

6 & Female & 20-27 & Chinese, American &   Several times. 3-10 times in a lifetime. \\

7 & Male & 20-27 & Chinese & Several times. 3-10 times in a lifetime. \\

8 & Male & 20-27 & Chinese, American & Several times. 3-10 times in a lifetime. \\

9 & Male & 20-27 & Chinese, Australian & Once or twice in a lifetime. \\

10 & Male & 20-27 & Chinese &   Never heard of    Peking Opera. \\

11 & Male & 20-27 & Chinese &   Never heard of    Peking Opera. \\

12 & Male & 20-27 & Chinese, American &   Never heard of    Peking Opera. \\

13 & Female & 20-27 & Chinese, American &   Never heard of    Peking Opera. \\

14 & Male & 28-35 & Chinese, American & Once or twice in a lifetime. \\

15 & Male & 28-35 & Chinese, American & Once or twice in a lifetime. \\

16 & Male & 35+ & Chinese, Canadian & Several times. 3-10 times in a lifetime. \\
\hline
\end{tabular}
\caption{Demographic Information of Phase 3 Study Participants (N=16)}
\label{tab:participant_info}
\end{table*}

\subsection{Procedure}
With intentionality, we limit the entire experience to less than half an hour\cite{Han2022PrerequisitesforLearninginvr}, with the total estimated VR usage time not exceeding 15 minutes. Moreover, we plan to allocate 10 minutes for onboarding activities, and less than 15 minutes for total duration of the temporal audio. The process began with obtaining informed consent from each participant. participants were informed that they could withdraw from the study at any time for any reason without penalty. Participants were reminded that they could pause or stop the experience at any point if they felt uncomfortable or wished to take a break. Participants were equipped with a VR headset (Quest 2) and given instructions on its proper use. They were directed to enter our prototype URL in the VR browser to access the experience. Upon opening the VR scene, participants were instructed to wait 10 seconds for the environment to load before starting their exploration. Participants were then guided to click the VR icon in the bottom right corner of the browser tab to switch from WebVR mode to full VR mode. They were taught to interact with audio points by directing the controller at them, triggering automatic playback without clicking. The color-coding system was explained. Throughout the experience, participants were encouraged to think aloud, providing real-time verbal feedback. Participants were informed they could exit a scene at any time. After completing each scene, they were instructed to press the circle button on the right-hand controller to return to the browser version and select "Download Trajectory." This process was repeated for all four scenes. The study concluded with a semi-structured interview.

\subsection{Data Collection and Analysis}
Our study utilized semi-structured interviews, think-aloud protocols\cite{Charters2003_thinkaload}, and observations as the main data sources, supplemented by system logs to provide additional supportive validation. 
During the VR experiences, we employed the think-aloud method\cite{Charters2003_thinkaload}, where participants provided real-time verbal feedback as they engaged with the four VR scenarios. This approach offered immediate insights into participants' thought processes and reactions, capturing their unfiltered responses to the virtual environments.
To further enrich our qualitative data, we used a phone to record videos of the participants' physical interactions, their think-aloud protocol\cite{Charters2003_thinkaload}, and any spontaneous comments. These observations provided valuable visual data on how participants physically engaged with the VR technology and responded to the cultural content. The interviews gathered in-depth insights into participants' experiences, preferences, and comparisons with other audiovisual media transmission methods, forming the primary basis for our analysis. \cite{BraunClarke2006_thematicanalysis, Lungu2022_QUALcoding} of interview transcripts and think-aloud\cite{Charters2003_thinkaload} data was conducted to uncover insights into user experiences, exploration patterns, impact of spatial interactions on cultural understanding, and comparisons with other media transmission methods. The transcripts of the interviews and think-aloud data were coded by two researchers separately \cite{BraunClarke2006_thematicanalysis, Lungu2022_QUALcoding} . The two researchers then revised the codes and discussed these themes with the entire research team. This process was repeated iteratively until final consensus was reached among all team members. The observations allowed us to directly observe audience behaviors and interactions in real-time.
During these sessions, we systematically captured field notes. All observational data were collected and organised on Miro, for visualising and analysing patterns in audience interaction patterns. Lastly, we captured trajectory data through system logs, which recorded user interactions within the VR environment, including dwell time at each audio segment and overall experience duration, and sequence of audio point IDs that participants interacted with. These information provided objective measures to support and contextualize our primary findings.

\subsection{Phase 3 Study Results: User Interaction Patterns}
Our analysis of 16 participants revealed rich insights into how users navigate and make sense of spatially-arranged ICH audio content in SISA. 


\subsubsection{Progressive Exploration Pattern}

\begin{figure*} [ht]
    \centering
    \includegraphics[width=1\linewidth]{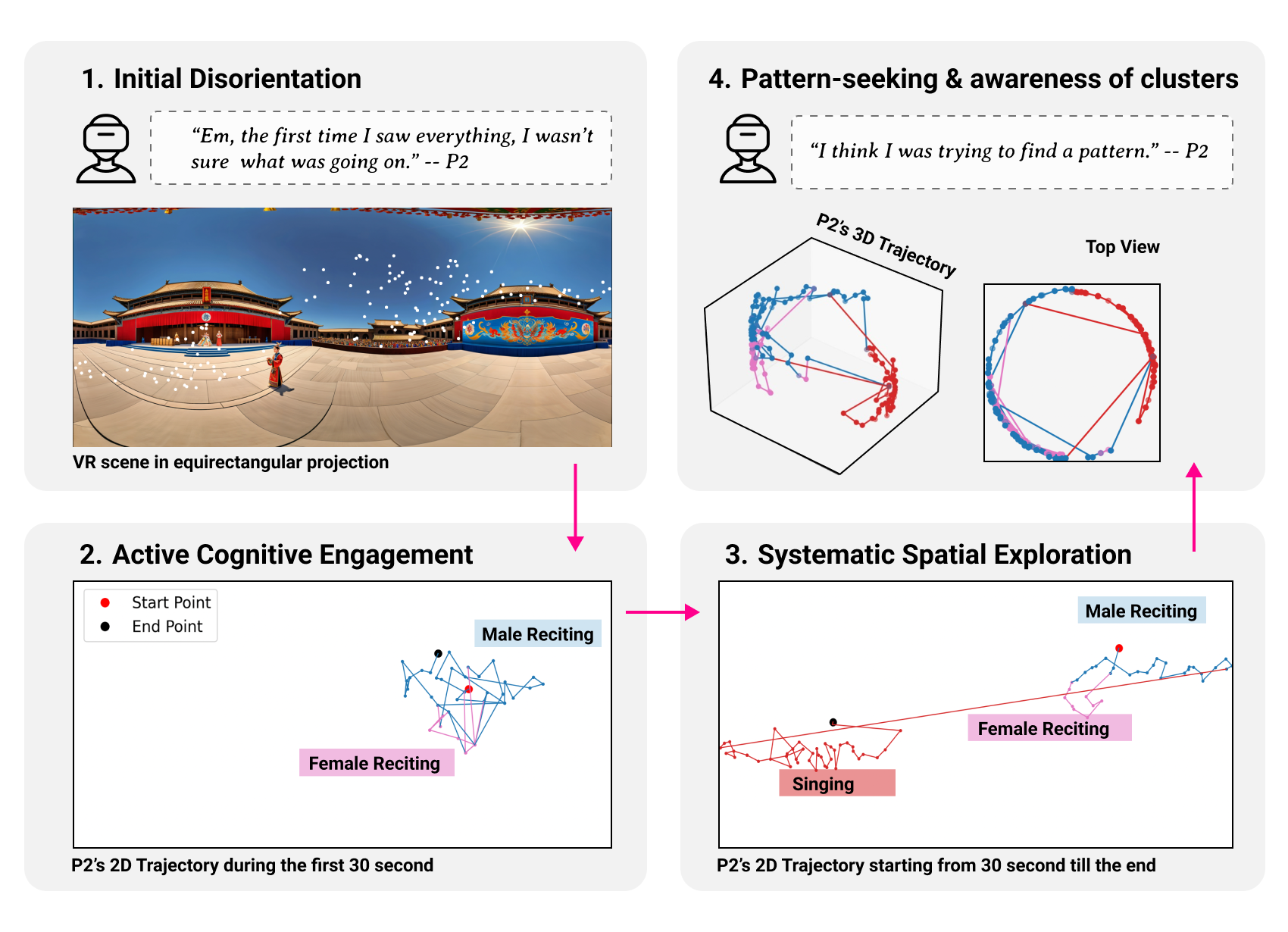}
    \caption{Progressive Exploration Pattern. The sequence illustrates Participant 2’s transformation from 1) initial disorientation through 2) active cognitive engagement to develop 3) a systematic spatial exploration to 4) pattern-seeking and awareness of spatial audio clusters.  (\textbf{Top Left}) Equirectangular Projection of the Peking Opera VR scene. (\textbf{Bottom Left}) The 2D trajectory of Participant 2's engagement with audio points during the first 30 seconds, showing disoriented exploration between the two reciting clusters. (\textbf{Bottom Right}) The trajectory of Participant 2’s remaining experience after the initial 30 seconds. P2 has shown patterns to adopt a more systematic exploring between clusters.(\textbf{Top Right}) The 3D trajectory and top view illustrate how Participant 2 identified audio clusters, with transitions between singing and reciting clusters in the 360-degree spatial environment, indicating an emerging understanding of their distinctions. }
    \label{fig:progressive}
\end{figure*}

Our prototype offers a new approach to listening to auditory ICH, requiring participants spatially interacting with auditory ICH segments in the immersive environment. Participants all reported their exploration advanced from  an 1) initial disorientation through 2) active cognitive engagement to develop 3) a systematic exploration to 4) pattern-seeking and awareness of spatial audio clusters, transforming traditional passive auditory ICH content consumption into a more active and engaging experience. This progression is illustrated by an example shown in Figure \ref{fig:progressive}. \add{This progression from disorientation to pattern recognition suggests that SISA facilitates an understanding of the genre’s musical structure through spatial exploration. Users gradually identify distinctive genre elements—such as recitation, singing, or character-specific sonic signatures — that would typically require extensive listening to discern.}

\paragraph{1) Initial Disorientation} 
Participant 2 initially experienced disorientation when encountering the SISA environment, and commented \textit{"Em, the first time I saw everything, I wasn’t sure  what was going on"}. This disorientation stemmed from the unfamiliarity of the spatial representation of traditionally temporal music content and was attributed to participants' accustomed expectations of traditional temporal forms of music content consumption. Our prototype's design, which segmented temporal music into spatial audio segments, challenged the audience's typical passive content consumption style.
\paragraph{2) Active Cognitive Engagement}
Despite initial confusion, as participants progress, this new exploration method triggers their curiosity, motivating them to actively engage and explore different points, seeking to understand the relationships between the spatially arranged audio segments. This indicates that the spatial arrangement encouraged analytical thinking and deeper engagement with the ICH material. Participant 5 remarked on the perceived intentionality of the design:
\textit{"I think these design elements may have a certain distribution. I guess this distribution might be intended to guide me to watch and listen simultaneously."} These responses indicate that the spatial-audio environment successfully engages users to move beyond passive listening to active exploration, interaction and interpretation of the auditory ICH content.
\paragraph{3) Systematic Spatial Exploration}
To navigate the unfamiliar environment, participants developed systematic spatial exploration strategies. Participant 2 described their approach:
\textit{"It was a 360-degree picture, so I just went in a circle between different altitudes and usually checked the top as well to make sure I saw everything."}
Other participants (P4,14,15) reported similar methodical approaches, such as \textit{"First exploring the audio segments in front of me, then listening to the nearby audio segments around this segment."}
Participant 14 elaborated further:
\textit{"Then I would look at the audio segments further away to see if they were very different from the closer segments."}
These strategies demonstrate that participants actively tried to understand the relationships between spatially arranged audio segments while exploring the meaning of audio segments placed far apart.
\paragraph{4) Pattern-seeking and Awareness of Spatial Audio Clusters.}
As participants continued their exploration, many began to engage in pattern-seeking behavior, attempting to discern intentional relationships between the spatially arranged audio segments. For instance, Participant 2 reflected on their experience: \textit{"I think I was trying to find a pattern."}
This indicates that the spatial arrangement encouraged analytical thinking and deeper engagement with the ICH material. Participant 5 remarked on the perceived intentionality of the design:
\textit{"I think these design elements may have a certain distribution. I guess this distribution might be intended to guide me to watch and listen simultaneously."} These responses indicate that the spatial-audio environment successfully engages users to move beyond passive listening to active exploration, interaction and interpretation of the auditory ICH content.

\subsubsection{Adaptive Exploration Strategy: Transitioning from Visual to Audio-Driven.}
\begin{figure*}[ht]
    \centering
    \includegraphics[width=0.8 \linewidth]{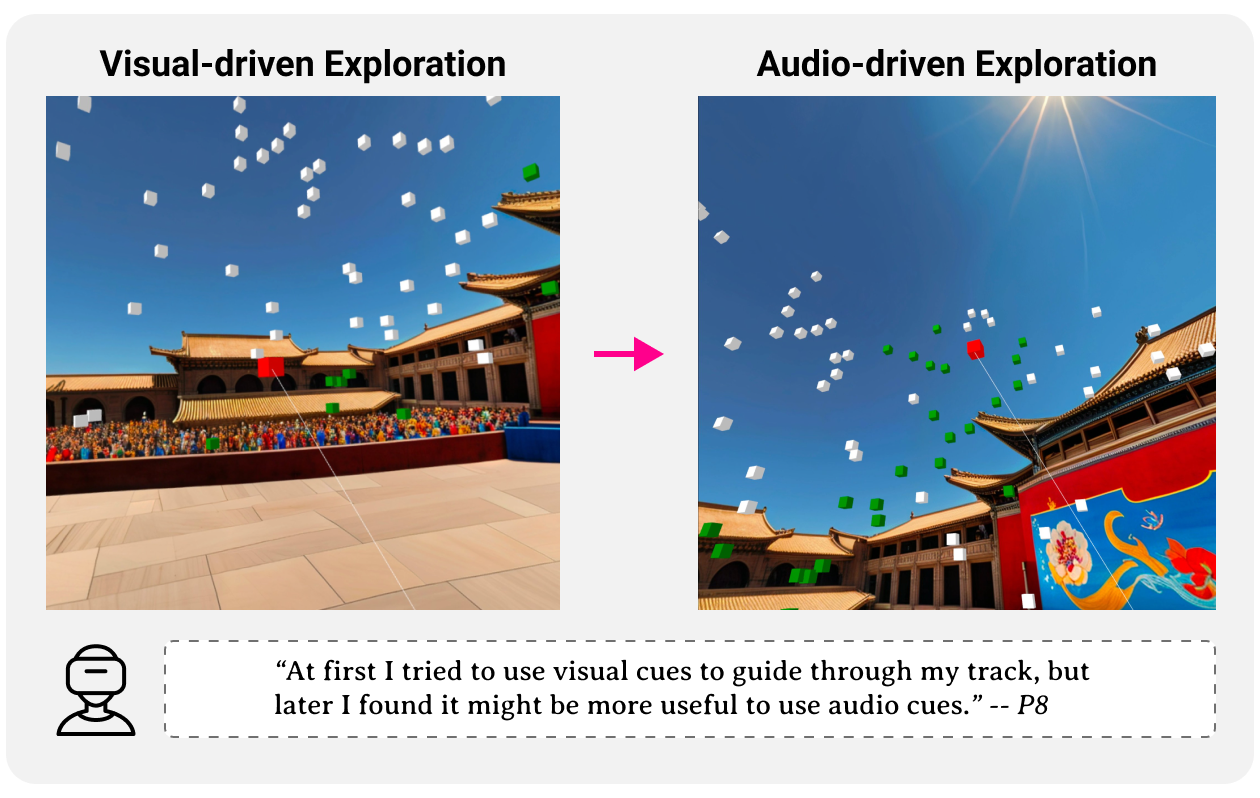}
    \caption{Adaptive exploration strategy: transition from visual-driven to audio-driven exploration in the SISA environment. 
   \textbf{ (Left)}: Initial visual-driven interaction in the VR experience. 
    \textbf{(Right)} Evolved and adapted audio-driven approach, where Participant 8 identified audio clusters independent of visual cues. This shift, exemplified by P8's comment, demonstrates the participants' adaptation to SISA environment.}
    \label{fig:adaptivefinal}
\end{figure*}

Participants initially relied on visual cues to navigate the spatial audio environment but gradually shifted towards using audio cues as their primary guide. This transition, illustrated in Figure \ref{fig:adaptivefinal}, marked a significant change in their exploration strategy and engagement with the content. Participant 8 clearly articulated this shift: \textit{"At first I tried to use visual cues to guide through my track, but later I found it might be more useful to use audio cues."} 
This transition from visual to audio-driven exploration was also evident in other participants' behaviors. For example, Participant 4's observations marked a shift towards more focused auditory recognition: \textit{"This area is all musical bangu sounds. That area is all reciting."} This comment illustrates how participants began distinguishing between different sound clusters, setting the stage for more nuanced auditory analyses. Expanding further, Participant 14's analysis reflects a sophisticated engagement with the material: 
\textit{"I feel like these points might be from the same part of an opera piece. Because the sound and singing style are similar. I tried to connect them, so I listened to all the surrounding ones."}
This shift from a primarily visual-oriented approach to one that is more driven by audio cues demonstrates the participants' adaptability and the design's effectiveness in emphasizing auditory over visual information, guiding their exploration and connection-building between segments.
The shift from visual to audio-driven exploration also came with a learning curve, as noted by Participant 15:
\textit{"But once you figure it out, it becomes pretty intuitive after that."}

\section{Discussion: Design Insights for SISA}\label{sec:Discussion}
Our findings 
highlighted promising opportunities where temporal-to-spatial SISA approach could enhance traditional auditory ICH engagement. 
\add{Through the lens of Schaeffer's four modes of listening \cite{Schaeffer_4listeningModes} — Écouter (listening to identify sources), Ouïr (passive perception), Entendre (selective hearing), and Comprendre (comprehending meaning) — we can interpret how the observed Progressive and Adaptive exploration patterns suggest potential pathways through increasingly sophisticated levels of auditory engagement with ICH content.} In this section, we synthesize the design insights distilled from our findings. These insights reflect a combination of design considerations identified through expert evaluation, interaction patterns that emerged from our user study, and broader implications for the future of ICH preservation and engagement enhanced by machine learning algorithms and immersive technologies. They are intended to provide actionable guidance for designers and researchers working on interactive system design and development for ICH engagement and preservation, particularly in the context of auditory ICH such as performing arts.

\subsection{Design Insight 1: Re-frame auditory ICH engagement, a temporal-to-spatial approach }

Our research demonstrates how the SISA prototype fundamentally transforms traditional temporal-based auditory ICH experiences into spatially-oriented audio-driven interactive engagements. This transformation, anchored in our modular presentation principle (DR-1) and accessibility enhancement objectives (DC-2), also addresses critical ICH sustainability challenges by decoupling cultural engagement from location constraints. \add{The temporal-to-spatial transformation enables accelerated comprehension of genre-specific characteristics through interactive exploration. By spatially clustering similar audio properties, SISA creates a navigable taxonomy that reveals essential patterns and variations defining Peking Opera. This approach allows even novices to distinguish between character types, vocal techniques, and emotional qualities through progressive engagement levels, while preserving cultural authenticity.} Phase 3 findings identified a distinct progressive exploration pattern that aligns with Chi and Wylie's ICAP framework \cite{Chi2014The}. Users typically advanced from initial spatial disorientation through increasingly sophisticated engagement levels: from passive observation to active exploration, constructive pattern recognition, and ultimately interactive meaning-making through audio cluster interpretation. This progression provides empirical validation of the ICAP hypothesis, which posits enhanced learning outcomes through hierarchical engagement advancement \cite{Chi2014The,Qin2023EXPRESS:}.


\subsection{Design Insight 2: Enhance engagement with auditory ICH, from visual to audio-driven}

Our SISA prototype introduces an approach that transforms auditory ICH into interactive spatial experiences, advancing beyond traditional VR applications that primarily focus on visual elements \cite{Cai2023The, Zhang2023A}. Our work provides designers with clear design rationale and steps for converting temporal-based auditory ICH content into engaging spatially interactive experiences.

The adaptive exploration pattern emerged from our phase 3 user study demonstrate users' natural transition from visual-driven to audio-driven exploration strategy, validating the effectiveness of our temporal-to-spatial transformation approach. This adaptivon supports our scaffolded exposure approach (DR-2) while aligning with UNESCO's emphasis \cite{unesco2003basictexts} on active participation in cultural preservation. The framework extends the HCI toolkit for designing engaging interactive systems for ICH preservation, offering insights that can be adapted across various cultural contexts.

Both adaptive and progressive exploration patterns from our findings directly address documented limitations of passive reception in traditional auditory ICH experiences \cite{rao2024formationcreator, Liu_Hua'er, lu2011shadowstory, yao2024shadowmaker, Othman2021Overview}. Through prioritizing auditory over visual engagement and facilitating spatial exploration, we establish a novel approach grounded in modular presentation (DR-1) that enhances accessibility (DC-2) while providing scaffolded exposure (DR-2) to auditory ICH."


\subsection{Design Insight 3: Balance exploration and guidance}
Our analysis demonstrates the importance of balanced navigation systems in ICH interaction design. These systems must support unstructured exploration while integrating strategic guidance mechanisms. Based on empirical observations from Phase 2 and 3 studies, we identified that effective ICH engagement requires two key elements: support for natural exploratory behaviors and guidance toward essential cultural elements, addressing our design consideration for scaffolded exposure (DC-2). 

The implementation strategies should combine visual and auditory guidance, aligning with our multi-sensory design rationale (DR-3). In the SISA final prototype, visual feedback is provided through color-coded audio segments, differentiating between explored and unexplored cultural components. For auditory guidance, our SISA approach implements spatially-clustered audio segments that direct users toward significant modular elements (DR-1) of the selected ICH. This integrated multi-sensory approach (DR-3) facilitates a progressive revelation of content that corresponds to users' exploration patterns.


\subsection{Design Insight 4: Balance engagement and learning}
Our methodology, validated through observed progressive exploration patterns, establishes a systematic approach to embedding contextual information within immersive environments. Systems should integrate modular presentation (DR-1) of cultural knowledge, creating interconnected learning pathways that support various engagement levels while preserving cultural authenticity (DC-1). The effectiveness of this approach is evidenced by participants' successful transition from initial exploration to deeper engagement with cultural content, as observed in our Phase 3 study.


\subsection{Design Insight 5: Balance technology and authenticity}
Our observation reveals that the current time-based segmentation approach (5 or 10 seconds) in SISA prototypes can be sometimes disruptive to the natural flow of some performances, suggesting the necessity for more culturally informed segmentation methodologies. \remove{We propose that future implementations should prioritize segmentation based on the "internal essence" (DR-1) of performances, specifically incorporating natural boundaries defined by structural elements such as lyrical phrases and traditional performance segments.} \add{While time-based segmentation provides technical consistency, it occasionally fragments natural performance structures. We propose that future implementations should prioritize segmentation based on the "internal essence" (DR-1) of performances—specifically aligning with Peking Opera’s inherent architecture by respecting natural boundaries defined by lyrical phrases, rhythmic patterns, and dramatic transitions. This culturally sensitive approach would preserve semantic integrity while enhancing accessibility through modular presentation. Its implementation requires extensive collaboration with ICH practitioners to identify meaningful content boundaries that maintain the traditional performance experience while minimizing disruptive interventions.} 
The implementation of such culturally sensitive segmentation methods requires extensive collaboration with ICH practitioners to establish meaningful content boundaries that preserve traditional performance elements while minimizing disruptive interventions in the cultural experience.

\subsection{Design Insight 6: Opportunities for fine-tuning t-SNE algorithms in auditory ICH preservation and engagement}
Our work extends the application of t-SNE algorithms into a novel domain: the clustering of auditory ICH content to facilitate interactive spatial engagement experiences. The efficacy of our audio-clustering approach is validated through empirical observation of users successfully identifying and navigating distinct audio clusters, suggesting promising avenues for developing technologically enhanced preservation methods that maintain cultural authenticity (DC-1) while improving accessibility (DC-2). As an early attempt, we focus more on investigating how people listening to and interact with the SISA prototypes rather than developing a systematic method for fine-tuning the t-SNE algorithm specifically for ICH content. 
Future research should focus on developing systematic methods for optimizing clustering algorithms specifically for ICH content while maintaining the balance between technological innovation and cultural preservation. 

\subsection{Contributions and Implications}
The above insights highlight the contributions of our work to auditory ICH interaction design. 
\remove{Our research advances the methodology for designing sustainable cultural heritage interactions through a rigorously structured three-phase process: participatory co-design, iterative prototyping, and comprehensive user testing.}\add{Our research advances understanding of the methodology for designing sustainable cultural heritage interactions through a structured three-phase process: participatory co-design, iterative prototyping, and comprehensive user testing.} 

The SISA system's development framework \remove{demonstrates the efficacy of} \add{suggests the potential of} temporal-to-spatial transformation in cultural heritage preservation, \remove{establishing a replicable methodology that reconceptualizes how temporal performances can be spatially preserved and experienced} \add{offering a potentially replicable methodology that reconceptualizes how temporal performances might be spatially preserved and experienced}. \remove{This novel interaction paradigm extends beyond conventional user-centered design approaches by introducing spatially-interactive digital environments that reduce resource dependencies while maintaining cultural authenticity.}

Our work contributes to the discourse on sustainable digital heritage preservation. The transformation of traditionally resource-intensive cultural experiences into digital environments provides \add{insights into} an environmentally conscious framework while \remove{ensuring} \add{seeking to ensure} cultural authenticity. The methodological framework offers researchers and practitioners \remove{with empirically validated} protocols for developing and implementing interactive systems in the ICH domain. This systematic approach establishes a foundation for methodological consistency and \remove{positions interactive systems} \add{explores the role of interactive systems} as integral components in supporting the long-term resilience of ICH preservation initiatives.

\subsection{Limitations}

\remove{Our study faced several key limitations. Sample limitations impact the generalizability of our results. Our user testing involved 16 participants, predominantly with Chinese cultural backgrounds, which may not fully represent the broader potential ICH audience. The focus on only one ICH genres, Peking Opera, further limits the scope. Technical limitations were evident in our audio segmentation approach. The time-based divisions (5 or 10 seconds) can be disruptive to the natural flow of performances, indicating a need for more culturally sensitive segmentation methods. Inconsistent total exploration time among participants introduced variability in exposure and engagement with the SISA prototype.} 

\add{
\subsubsection{Cultural genre Limitations.}
Our focus on Peking Opera constrains generalizability, embedding insights within Chinese cultural contexts. This specificity affects audio segmentation and clustering functionality, as different traditions employ varying rhythmic structures and tonal patterns requiring different clustering approaches. The t-SNE algorithm may perform differently when applied to other genres such as Indian classical music's tala patterns or African polyrhythmic traditions.}
\add{
\subsubsection{Sampling and Representation Challenges.}
Our participant sample (n=16), predominantly individuals with Chinese cultural backgrounds (11 mixed, 5 solely Chinese), limits the analytical scope for broader SISA framework applications. This culturally homogeneous sampling, while appropriate for our Peking Opera focus, creates a specific interpretive lens that may not account for how culturally unfamiliar audiences would navigate and make sense of spatially transformed ICH content. This limitation is particularly significant given that our design rationale explicitly included accessibility enhancement (DC-2) to address how to "help people who don't speak Chinese understand the deeper value of our art, including its cultural significance and abstract elements" (CoD3). The absence of participants without Chinese cultural knowledge means we cannot fully evaluate how effectively our approach bridges cultural divides—a critical consideration for ICH preservation and transmission in an increasingly global context.}
\add{
\subsubsection{Algorithmic Parameter Exploration.}
Our implementation of the t-SNE algorithm for audio clustering, while demonstrating a proof of concept, represents a constrained exploration of the potential parameter space. The efficacy of t-SNE is highly dependent on parameters such as perplexity and learning rate, which significantly influence the resulting spatial organization of audio segments. As noted in our methodology, we selected parameters based on observed outcomes where "minimum similarity across segments < 0.9 and KL divergence < 1, ideally < 0.5," but did not systematically investigate how different parameter configurations might affect user experiences and engagement patterns. This analytical gap limits our understanding of how algorithm optimization might enhance or detract from cultural authenticity (DC-1), particularly when considering the balance between technological innovation and preservation of a genre's "internal essence" (DR-1).}
\add{
\subsubsection{Visual Environment Generation limitation.}
While our AI generated environment creation approach aims to produce contextually relevant visual scenes, we recognize the inherent limitations of AI in representing authentic cultural spaces. The generated environments, though visually compelling, cannot fully capture the architectural nuances, historical accuracy, and atmospheric qualities of traditional performance venues. The visual fidelity of AI-generated environments and their alignment with genre characteristics were not assessed in this paper, as it falls outside the scope of our study. The AI-generated imagery raises questions about authenticity that align with our design consideration for cultural authenticity preservation (DC-1). As echoed by CoD3, traditional opera serves as "a living fossil of our culture," suggesting that visual representation requires advanced examination and preservation care.}

\add{
\subsubsection{Methodological Study Limitations.}
Our study design brings some methodological constraints affecting interpretation of SISA prototype interactions. The inconsistent exploration time among participants — resulting from our naturalistic, self-guided approach — created variability in engagement levels that complicates comparative analysis and potentially obscures patterns in spatial audio comprehension development. Our reliance on think-aloud protocols, while yielding rich qualitative data, likely altered participants' natural exploration patterns and introduced artificial reflective processes. The absence of control groups comparing traditional linear listening with our spatial approach further limits quantitative assessment of spatial transformation's impact on engagement and learning outcomes.}

\add{
\subsection{Implications}
The SISA approach demonstrates significant potential for adaptation across various ICH dissemination contexts beyond immersive environments. The temporal-to-spatial transformation methodology established in this study offers a conceptual foundation that could enhance other interactive media platforms while addressing UNESCO's call for making ICH "truly alive" \cite{unesco2003basictexts} 

Our SISA approach could be effectively integrated into serious games, where players navigate cultural soundscapes as part of gameplay mechanics. Similar to how Ch'ng et al. demonstrated improved learning outcomes through gamified cultural heritage experiences \cite{Ch’ng2020The}, SISA design insights could transform passive musical content into interactive game elements where progression depends on recognizing and categorizing traditional performance features. For cinematic and documentary presentations, the SISA clustering methodology could inform interactive film interfaces where viewers navigate between narrative segments based on acoustic similarities, creating personalized engagement pathways analogous to Lu et al.'s implementation of livestreaming for cultural transmission \cite{lu2019feel}.

The SISA system's practical deployment would be particularly effective in museum contexts as "interactive soundscape stations." Visitors could use lightweight VR headsets at designated exhibition areas to explore traditional Peking Opera elements through spatial navigation, with each station focusing on specific performance aspects (vocal techniques, instrumental sections, character types). Similar to Tsita et al.'s VR museum implementation \cite{Tsita2023A}, these stations would function as self-contained exploration pods requiring minimal staff supervision while providing rich engagement opportunities. The interface would offer graduated interaction levels—from guided exploration for novices to free navigation for experienced users—addressing the accessibility enhancement considerations (DC-2) identified in our design framework. The experience resembles "tuning on the radio" but with spatial dimensions, creating an intuitive entry point for unfamiliar cultural content.}

\subsection{Future Work}
These limitations provide valuable directions for future research, including the development of content-aware segmentation algorithms, creation of validated ICH engagement measurement tools, and conducting longitudinal studies \add{with larger, more diverse samples. Future iterations could also incorporate eye-tracking to precisely map attention patterns during the transition from disorientation to engagement. Additionally,} future studies should consider including control groups and standardizing exploration times to more robustly assess the SISA prototype's impact. 

Recently, Generative AI (GenAI) has been applied to creative applications of storytelling\cite{yang_ai_2022}, gamefied interactions\cite{zhang_can_2025, ling_sketchar_2024}, creative design\cite{han_when_2024}, and intangible cultural heritage narratives\cite{liu_salt_2025}. The power of Large Language Models in the heritage recording process\cite{zhou_retrochat_2025} is tempered by their inherent biases and negative use tendencies\cite{zeng_ronaldos_2025}. Future work can examine how purposive storytelling approaches\cite{lc_designing_2021, lc_designing_2022, zhou_eternagram_2024, lc_case_2022} may be enabled by GenAI for fair and just representations of ICH in the technological performance space\cite{lc_contradiction_2023, lc_active_2023}.

\section{Conclusion}\label{sec:Conclusion}

The SISA approach \remove{reimagines} \add{presents a novel perspective on} auditory ICH engagement by transforming temporal listening into active spatial interaction within immersive VR environments. Through a structured, three-phase research process—co-designing with stakeholders, iterative prototyping, and user testing—we identified critical design elements and \remove{validated the efficacy of} \add{explored the potential of} spatial audio exploration. Our findings \remove{highlight} \add{suggest} SISA's \remove{efficacy} \add{potential} to foster engagement, particularly with at-risk genres like Peking Opera. By addressing specific challenges in ICH engagement and contributing to discussion around UNESCO's vision of keeping ICH vibrant, this study \remove{establishes} \add{offers insights toward} a methodological foundation for future exploration of ICH interactive systems. 




\begin{thebibliography}{90}


\ifx \showCODEN    \undefined \def \showCODEN     #1{\unskip}     \fi
\ifx \showDOI      \undefined \def \showDOI       #1{#1}\fi
\ifx \showISBNx    \undefined \def \showISBNx     #1{\unskip}     \fi
\ifx \showISBNxiii \undefined \def \showISBNxiii  #1{\unskip}     \fi
\ifx \showISSN     \undefined \def \showISSN      #1{\unskip}     \fi
\ifx \showLCCN     \undefined \def \showLCCN      #1{\unskip}     \fi
\ifx \shownote     \undefined \def \shownote      #1{#1}          \fi
\ifx \showarticletitle \undefined \def \showarticletitle #1{#1}   \fi
\ifx \showURL      \undefined \def \showURL       {\relax}        \fi
\providecommand\bibfield[2]{#2}
\providecommand\bibinfo[2]{#2}
\providecommand\natexlab[1]{#1}
\providecommand\showeprint[2][]{arXiv:#2}

\bibitem[A-Frame(2021)]%
        {a-frame_2021}
\bibfield{author}{\bibinfo{person}{A-Frame}.} \bibinfo{year}{2021}\natexlab{}.
\newblock \bibinfo{title}{A-Frame}.
\newblock
\newblock
\urldef\tempurl%
\url{https://github.com/aframevr/aframe}
\showURL{%
\tempurl}


\bibitem[Abraham et~al\mbox{.}(2014)]%
        {abraham2014machine}
\bibfield{author}{\bibinfo{person}{Alexandre Abraham}, \bibinfo{person}{Fabian Pedregosa}, \bibinfo{person}{Michael Eickenberg}, \bibinfo{person}{Philippe Gervais}, \bibinfo{person}{Andreas Mueller}, \bibinfo{person}{Jean Kossaifi}, \bibinfo{person}{Alexandre Gramfort}, \bibinfo{person}{Bertrand Thirion}, {and} \bibinfo{person}{Ga{\"e}l Varoquaux}.} \bibinfo{year}{2014}\natexlab{}.
\newblock \showarticletitle{Machine learning for neuroimaging with scikit-learn}.
\newblock \bibinfo{journal}{\emph{Frontiers in neuroinformatics}}  \bibinfo{volume}{8} (\bibinfo{year}{2014}), \bibinfo{pages}{71792}.
\newblock


\bibitem[Arendttorp et~al\mbox{.}(2023)]%
        {arendttorp_ICH_VRgesture}
\bibfield{author}{\bibinfo{person}{Emilie Maria~Nybo Arendttorp}, \bibinfo{person}{Heike Winschiers-Theophilus}, \bibinfo{person}{Kasper Rodil}, \bibinfo{person}{Freja B.~K. Johansen}, \bibinfo{person}{Mads Rosengreen~J\o{}rgensen}, \bibinfo{person}{Thomas K.~K. Kjeldsen}, {and} \bibinfo{person}{Samkao Magot}.} \bibinfo{year}{2023}\natexlab{}.
\newblock \showarticletitle{Grab It, While You Can: A VR Gesture Evaluation of a Co-Designed Traditional Narrative by Indigenous People}. In \bibinfo{booktitle}{\emph{Proceedings of the 2023 CHI Conference on Human Factors in Computing Systems}} (Hamburg, Germany) \emph{(\bibinfo{series}{CHI '23})}. \bibinfo{publisher}{Association for Computing Machinery}, \bibinfo{address}{New York, NY, USA}, Article \bibinfo{articleno}{308}, \bibinfo{numpages}{13}~pages.
\newblock
\showISBNx{9781450394215}
\urldef\tempurl%
\url{https://doi.org/10.1145/3544548.3580894}
\showDOI{\tempurl}


\bibitem[Badrajan and Bunea(2023)]%
        {Badrajan2023Safeguarding}
\bibfield{author}{\bibinfo{person}{Svetlana Badrajan} {and} \bibinfo{person}{Diana Bunea}.} \bibinfo{year}{2023}\natexlab{}.
\newblock \showarticletitle{Safeguarding and researching the intangible musical heritage in the context of contemporary digital technologies}.
\newblock \bibinfo{journal}{\emph{Valorificarea și conservarea prin digitizare a colecțiilor de muzică academică și tradițională din Republica Moldova}} (\bibinfo{year}{2023}).
\newblock
\urldef\tempurl%
\url{https://doi.org/10.55383/digimuz2023.02}
\showDOI{\tempurl}


\bibitem[Bain(1969)]%
        {Bain1969Performing}
\bibfield{author}{\bibinfo{person}{Wilfred~C. Bain}.} \bibinfo{year}{1969}\natexlab{}.
\newblock \showarticletitle{Performing Arts: The Economic Dilemma}.
\newblock \bibinfo{journal}{\emph{Journal of Research in Music Education}} \bibinfo{volume}{17}, \bibinfo{number}{1} (\bibinfo{year}{1969}), \bibinfo{pages}{170--172}.
\newblock
\urldef\tempurl%
\url{https://doi.org/10.2307/3344206}
\showDOI{\tempurl}
\showeprint{https://doi.org/10.2307/3344206}


\bibitem[Beritelli and Grasso(2008)]%
        {4813723}
\bibfield{author}{\bibinfo{person}{Francesco Beritelli} {and} \bibinfo{person}{Rosario Grasso}.} \bibinfo{year}{2008}\natexlab{}.
\newblock \showarticletitle{A pattern recognition system for environmental sound classification based on MFCCs and neural networks}. In \bibinfo{booktitle}{\emph{2008 2nd International Conference on Signal Processing and Communication Systems}}. \bibinfo{pages}{1--4}.
\newblock
\urldef\tempurl%
\url{https://doi.org/10.1109/ICSPCS.2008.4813723}
\showDOI{\tempurl}


\bibitem[Braun and Clarke(2006)]%
        {BraunClarke2006_thematicanalysis}
\bibfield{author}{\bibinfo{person}{Virginia Braun} {and} \bibinfo{person}{Victoria Clarke}.} \bibinfo{year}{2006}\natexlab{}.
\newblock \showarticletitle{Using thematic analysis in psychology}.
\newblock \bibinfo{journal}{\emph{Qualitative Research in Psychology}} \bibinfo{volume}{3}, \bibinfo{number}{2} (\bibinfo{year}{2006}), \bibinfo{pages}{77--101}.
\newblock
\urldef\tempurl%
\url{https://doi.org/10.1191/1478088706qp063oa}
\showDOI{\tempurl}


\bibitem[Brown(2005)]%
        {Brown2005}
\bibfield{author}{\bibinfo{person}{Michael Brown}.} \bibinfo{year}{2005}\natexlab{}.
\newblock \showarticletitle{Heritage Trouble: Recent Work on the Protection of Intangible Cultural Property}.
\newblock \bibinfo{journal}{\emph{International Journal of Cultural Property}}  \bibinfo{volume}{12} (\bibinfo{date}{02} \bibinfo{year}{2005}), \bibinfo{pages}{40 -- 61}.
\newblock
\urldef\tempurl%
\url{https://doi.org/10.1017/S0940739105050010}
\showDOI{\tempurl}


\bibitem[Burla and Yadav(2022)]%
        {Burla2022REVALUATION}
\bibfield{author}{\bibinfo{person}{Venkata~Naresh Burla} {and} \bibinfo{person}{Sudarshan Yadav}.} \bibinfo{year}{2022}\natexlab{}.
\newblock \showarticletitle{REVALUATION OF TRADITIONAL PERFORMING ARTS IN THE POST-INDEPENDENT INDIAN THEATRE}.
\newblock \bibinfo{journal}{\emph{ShodhKosh: Journal of Visual and Performing Arts}} (\bibinfo{year}{2022}).
\newblock
\urldef\tempurl%
\url{https://doi.org/10.29121/shodhkosh.v3.i2.2022.180}
\showDOI{\tempurl}


\bibitem[Cai and Yang(2023)]%
        {Cai2023The}
\bibfield{author}{\bibinfo{person}{Yunyan Cai} {and} \bibinfo{person}{Cao Yang}.} \bibinfo{year}{2023}\natexlab{}.
\newblock \showarticletitle{The Application and Research of VR Animation Technology in Intangible Cultural Heritage: –Take Danzhai Miao Batik as an example}.
\newblock \bibinfo{journal}{\emph{Proceedings of the 2023 8th International Conference on Information and Education Innovations}} (\bibinfo{year}{2023}).
\newblock
\urldef\tempurl%
\url{https://doi.org/10.1145/3594441.3594467}
\showDOI{\tempurl}


\bibitem[Cerquetti and Ferrara(2018)]%
        {CerquettiFerrara2018}
\bibfield{author}{\bibinfo{person}{Mara Cerquetti} {and} \bibinfo{person}{Concetta Ferrara}.} \bibinfo{year}{2018}\natexlab{}.
\newblock \showarticletitle{Marketing research for cultural heritage conservation and sustainability: Lessons from the field}.
\newblock \bibinfo{journal}{\emph{Sustainability}} \bibinfo{volume}{10}, \bibinfo{number}{3} (\bibinfo{year}{2018}), \bibinfo{pages}{774}.
\newblock


\bibitem[Charters(2003)]%
        {Charters2003_thinkaload}
\bibfield{author}{\bibinfo{person}{Elizabeth Charters}.} \bibinfo{year}{2003}\natexlab{}.
\newblock \showarticletitle{The Use of Think-aloud Methods in Qualitative Research An Introduction to Think-aloud Methods}.
\newblock \bibinfo{journal}{\emph{Brock Education Journal}} \bibinfo{volume}{12}, \bibinfo{number}{2} (\bibinfo{date}{July} \bibinfo{year}{2003}).
\newblock
\urldef\tempurl%
\url{https://doi.org/10.26522/brocked.v12i2.38}
\showDOI{\tempurl}


\bibitem[Chi and Wylie(2014)]%
        {Chi2014The}
\bibfield{author}{\bibinfo{person}{Michelene~TH Chi} {and} \bibinfo{person}{Ruth Wylie}.} \bibinfo{year}{2014}\natexlab{}.
\newblock \showarticletitle{The ICAP framework: Linking cognitive engagement to active learning outcomes}.
\newblock \bibinfo{journal}{\emph{Educational psychologist}} \bibinfo{volume}{49}, \bibinfo{number}{4} (\bibinfo{year}{2014}), \bibinfo{pages}{219--243}.
\newblock


\bibitem[Ch’ng et~al\mbox{.}(2020)]%
        {Ch’ng2020The}
\bibfield{author}{\bibinfo{person}{Eugene Ch’ng}, \bibinfo{person}{Yue Li}, \bibinfo{person}{Shengdan Cai}, {and} \bibinfo{person}{Fui-Theng Leow}.} \bibinfo{year}{2020}\natexlab{}.
\newblock \showarticletitle{The Effects of VR Environments on the Acceptance, Experience, and Expectations of Cultural Heritage Learning}.
\newblock \bibinfo{journal}{\emph{J. Comput. Cult. Herit.}} \bibinfo{volume}{13}, \bibinfo{number}{1}, Article \bibinfo{articleno}{7} (\bibinfo{date}{feb} \bibinfo{year}{2020}), \bibinfo{numpages}{21}~pages.
\newblock
\showISSN{1556-4673}
\urldef\tempurl%
\url{https://doi.org/10.1145/3352933}
\showDOI{\tempurl}


\bibitem[Cowen et~al\mbox{.}(2018)]%
        {cowen2018mapping}
\bibfield{author}{\bibinfo{person}{Alan~S Cowen}, \bibinfo{person}{Hillary~Anger Elfenbein}, \bibinfo{person}{Petri Laukka}, {and} \bibinfo{person}{Dacher Keltner}.} \bibinfo{year}{2018}\natexlab{}.
\newblock \showarticletitle{Mapping 24 Emotions Conveyed by Brief Human Vocalization}.
\newblock \bibinfo{journal}{\emph{American Psychologist}} \bibinfo{volume}{117}, \bibinfo{number}{4} (\bibinfo{year}{2018}), \bibinfo{pages}{1924--1934}.
\newblock
\urldef\tempurl%
\url{https://doi.org/10.1037/amp0000399}
\showDOI{\tempurl}


\bibitem[Cowen et~al\mbox{.}(2020)]%
        {cowen2020music}
\bibfield{author}{\bibinfo{person}{Alan~S Cowen}, \bibinfo{person}{Xia Fang}, \bibinfo{person}{Disa Sauter}, {and} \bibinfo{person}{Dacher Keltner}.} \bibinfo{year}{2020}\natexlab{}.
\newblock \showarticletitle{What music makes us feel: At least 13 dimensions organize subjective experiences associated with music across different cultures}.
\newblock \bibinfo{journal}{\emph{Proceedings of the National Academy of Sciences}} \bibinfo{volume}{117}, \bibinfo{number}{4} (\bibinfo{year}{2020}), \bibinfo{pages}{1924--1934}.
\newblock
\urldef\tempurl%
\url{https://doi.org/10.1073/pnas.1910704117}
\showDOI{\tempurl}


\bibitem[de~Miguel-Molina and Boix-Dom{\'e}nech(2021)]%
        {de-Miguel-Molina2021}
\bibfield{author}{\bibinfo{person}{Blanca de Miguel-Molina} {and} \bibinfo{person}{Rosina Boix-Dom{\'e}nech}.} \bibinfo{year}{2021}\natexlab{}.
\newblock \showarticletitle{Introduction: Music, from Intangible Cultural Heritage to the Music Industry}.
\newblock In \bibinfo{booktitle}{\emph{Music as Intangible Cultural Heritage}}, \bibfield{editor}{\bibinfo{person}{Blanca de~Miguel-Molina}, \bibinfo{person}{Victoria Santamarina-Campos}, \bibinfo{person}{Marina de~Miguel-Molina}, {and} \bibinfo{person}{Rosina Boix-Dom{\'e}nech}} (Eds.). \bibinfo{publisher}{Springer International Publishing}, \bibinfo{pages}{3--8}.
\newblock
\urldef\tempurl%
\url{https://doi.org/10.1007/978-3-030-76882-9_1}
\showDOI{\tempurl}


\bibitem[Dong et~al\mbox{.}(2014)]%
        {dong2014long}
\bibfield{author}{\bibinfo{person}{Li Dong}, \bibinfo{person}{Jiangping Kong}, {and} \bibinfo{person}{Johan Sundberg}.} \bibinfo{year}{2014}\natexlab{}.
\newblock \showarticletitle{Long-term-average spectrum characteristics of Kunqu Opera singers’ speaking, singing and stage speech}.
\newblock \bibinfo{journal}{\emph{Logopedics Phoniatrics Vocology}} \bibinfo{volume}{39}, \bibinfo{number}{2} (\bibinfo{year}{2014}), \bibinfo{pages}{72--80}.
\newblock


\bibitem[Dupont et~al\mbox{.}(2013)]%
        {dupont2013nonlinear}
\bibfield{author}{\bibinfo{person}{St{\'e}phane Dupont}, \bibinfo{person}{Thierry Ravet}, \bibinfo{person}{C{\'e}cile Picard-Limpens}, {and} \bibinfo{person}{Christian Frisson}.} \bibinfo{year}{2013}\natexlab{}.
\newblock \showarticletitle{Nonlinear dimensionality reduction approaches applied to music and textural sounds}. In \bibinfo{booktitle}{\emph{2013 IEEE International Conference on Multimedia and Expo (ICME)}}. IEEE, \bibinfo{pages}{1--6}.
\newblock


\bibitem[Erol et~al\mbox{.}(2022)]%
        {erol2022soundoff}
\bibfield{author}{\bibinfo{person}{Zeynep Erol}, \bibinfo{person}{Zhiyuan Zhang}, \bibinfo{person}{Eray {\"O}zg{\"u}nay}, {and} \bibinfo{person}{Ray LC}.} \bibinfo{year}{2022}\natexlab{}.
\newblock \showarticletitle{SOUND OF(F): Contextual Storytelling Using Machine Learning Representations of Sound and Music}. In \bibinfo{booktitle}{\emph{ArtsIT, Interactivity and Game Creation}}. Springer, \bibinfo{pages}{332--345}.
\newblock


\bibitem[Fedden(2017)]%
        {fedden_2017}
\bibfield{author}{\bibinfo{person}{Leon Fedden}.} \bibinfo{year}{2017}\natexlab{}.
\newblock \bibinfo{title}{Comparative Audio Analysis With Wavenet, MFCCs, UMAP, t-SNE and PCA}.
\newblock
\newblock
\urldef\tempurl%
\url{https://medium.com/@LeonFedden/comparative-audio-analysis-with-wavenet-mfccs-umap-t-sne-and-pca-cb8237bfce2f}
\showURL{%
\tempurl}


\bibitem[Fu et~al\mbox{.}(2023)]%
        {fu_i_2023}
\bibfield{author}{\bibinfo{person}{Kexue Fu}, \bibinfo{person}{Yixin Chen}, \bibinfo{person}{Jiaxun Cao}, \bibinfo{person}{Xin Tong}, {and} \bibinfo{person}{RAY LC}.} \bibinfo{year}{2023}\natexlab{}.
\newblock \showarticletitle{"{I} {Am} a {Mirror} {Dweller}": {Probing} the {Unique} {Strategies} {Users} {Take} to {Communicate} in the {Context} of {Mirrors} in {Social} {Virtual} {Reality}}. In \bibinfo{booktitle}{\emph{Proceedings of the 2023 {CHI} {Conference} on {Human} {Factors} in {Computing} {Systems}}} \emph{(\bibinfo{series}{{CHI} '23})}. \bibinfo{publisher}{Association for Computing Machinery}, \bibinfo{address}{New York, NY, USA}, \bibinfo{pages}{1--19}.
\newblock
\showISBNx{978-1-4503-9421-5}
\urldef\tempurl%
\url{https://doi.org/10.1145/3544548.3581464}
\showDOI{\tempurl}


\bibitem[Fu et~al\mbox{.}(2024)]%
        {fu_being_2024}
\bibfield{author}{\bibinfo{person}{Kexue Fu}, \bibinfo{person}{Ruishan Wu}, \bibinfo{person}{Yuying Tang}, \bibinfo{person}{Yixin Chen}, \bibinfo{person}{Bowen Liu}, {and} \bibinfo{person}{RAY LC}.} \bibinfo{year}{2024}\natexlab{}.
\newblock \showarticletitle{"{Being} {Eroded}, {Piece} by {Piece}": {Enhancing} {Engagement} and {Storytelling} in {Cultural} {Heritage} {Dissemination} by {Exhibiting} {GenAI} {Co}-{Creation} {Artifacts}}. In \bibinfo{booktitle}{\emph{Proceedings of the 2024 {ACM} {Designing} {Interactive} {Systems} {Conference}}} \emph{(\bibinfo{series}{{DIS} '24})}. \bibinfo{publisher}{Association for Computing Machinery}, \bibinfo{address}{New York, NY, USA}, \bibinfo{pages}{2833--2850}.
\newblock
\showISBNx{9798400705830}
\urldef\tempurl%
\url{https://doi.org/10.1145/3643834.3660711}
\showDOI{\tempurl}


\bibitem[Georgious Zaher~Georgious et~al\mbox{.}(2022)]%
        {georgious2022scalar}
\bibfield{author}{\bibinfo{person}{Ramy Georgious Zaher~Georgious}, \bibinfo{person}{K Elfeky}, \bibinfo{person}{RR Elrazky}, {et~al\mbox{.}}} \bibinfo{year}{2022}\natexlab{}.
\newblock \showarticletitle{Scalar, phasors and vector magnitudes for electric and electronic engineering}.
\newblock \bibinfo{journal}{\emph{Encyclopedia of Electrical and Electronic Power Engineering: Volumes 1-3}} (\bibinfo{year}{2022}).
\newblock


\bibitem[Han et~al\mbox{.}(2022)]%
        {Han2022PrerequisitesforLearninginvr}
\bibfield{author}{\bibinfo{person}{Eugy Han}, \bibinfo{person}{Kristine~L. Nowak}, {and} \bibinfo{person}{Jeremy~N. Bailenson}.} \bibinfo{year}{2022}\natexlab{}.
\newblock \showarticletitle{Prerequisites for Learning in Networked Immersive Virtual Reality}.
\newblock \bibinfo{journal}{\emph{Technology, Mind, and Behavior}} \bibinfo{volume}{3}, \bibinfo{number}{4} (\bibinfo{year}{2022}).
\newblock
\urldef\tempurl%
\url{https://doi.org/10.1037/tmb0000094}
\showDOI{\tempurl}


\bibitem[Han et~al\mbox{.}(2024)]%
        {han_when_2024}
\bibfield{author}{\bibinfo{person}{Yuanning Han}, \bibinfo{person}{Ziyi Qiu}, \bibinfo{person}{Jiale Cheng}, {and} \bibinfo{person}{RAY LC}.} \bibinfo{year}{2024}\natexlab{}.
\newblock \showarticletitle{When {Teams} {Embrace} {AI}: {Human} {Collaboration} {Strategies} in {Generative} {Prompting} in a {Creative} {Design} {Task}}. In \bibinfo{booktitle}{\emph{Proceedings of the {CHI} {Conference} on {Human} {Factors} in {Computing} {Systems}}} \emph{(\bibinfo{series}{{CHI} '24})}. \bibinfo{publisher}{Association for Computing Machinery}, \bibinfo{address}{New York, NY, USA}, \bibinfo{pages}{1--14}.
\newblock
\showISBNx{9798400703300}
\urldef\tempurl%
\url{https://doi.org/10.1145/3613904.3642133}
\showDOI{\tempurl}


\bibitem[He et~al\mbox{.}(2025)]%
        {he_i_2025}
\bibfield{author}{\bibinfo{person}{Zhiting He}, \bibinfo{person}{Jiayi Su}, \bibinfo{person}{Li Chen}, \bibinfo{person}{Tianqi Wang}, {and} \bibinfo{person}{RAY LC}.} \bibinfo{year}{2025}\natexlab{}.
\newblock \showarticletitle{"{I} {Recall} the {Past}": {Exploring} {How} {People} {Collaborate} with {Generative} {AI} to {Create} {Cultural} {Heritage} {Narratives}}.
\newblock \bibinfo{journal}{\emph{Proceedings of the ACM on Human-Computer Interaction}} \bibinfo{volume}{9}, \bibinfo{number}{CSCW 108} (\bibinfo{date}{April} \bibinfo{year}{2025}), \bibinfo{pages}{30}.
\newblock
\urldef\tempurl%
\url{https://doi.org/10.1145/3711006}
\showDOI{\tempurl}


\bibitem[{Hong Kong SAR Government}(2022)]%
        {hksarg2022pressrelease}
\bibfield{author}{\bibinfo{person}{{Hong Kong SAR Government}}.} \bibinfo{year}{2022}\natexlab{}.
\newblock \bibinfo{title}{Government announces appointment of Postmaster General}.
\newblock \bibinfo{howpublished}{Press Release}.
\newblock
\urldef\tempurl%
\url{https://www.info.gov.hk/gia/general/202211/09/P2022110900138.htm}
\showURL{%
\tempurl}


\bibitem[Hutchinson et~al\mbox{.}({[n.\,d.]})]%
        {hutchinsonTechnologyProbesInspiring2003}
\bibfield{author}{\bibinfo{person}{Hilary Hutchinson}, \bibinfo{person}{Wendy Mackay}, \bibinfo{person}{Bosse Westerlund}, \bibinfo{person}{Benjamin~B Bederson}, \bibinfo{person}{Allison Druin}, \bibinfo{person}{Catherine Plaisant}, \bibinfo{person}{Michel Beaudouin-Lafon}, \bibinfo{person}{Stéphane Conversy}, \bibinfo{person}{Helen Evans}, \bibinfo{person}{Heiko Hansen}, \bibinfo{person}{Nicolas Roussel}, \bibinfo{person}{Björn Eiderbäck}, \bibinfo{person}{Sinna Lindquist}, {and} \bibinfo{person}{Yngve Sundblad}.} \bibinfo{year}{[n.\,d.]}\natexlab{}.
\newblock \showarticletitle{Technology {{Probes}}: {{Inspiring Design}} for and with {{Families}}}.
\newblock  (\bibinfo{year}{[n.\,d.]}).
\newblock


\bibitem[Jameson(2021)]%
        {jameson_2021}
\bibfield{author}{\bibinfo{person}{Ben Jameson}.} \bibinfo{year}{2021}\natexlab{}.
\newblock \bibinfo{title}{SongDonkey.AI}.
\newblock
\newblock
\urldef\tempurl%
\url{https://songdonkey.ai/}
\showURL{%
\tempurl}


\bibitem[Kanagala and Jaya Rama~Krishnaiah(2016)]%
        {kanagala_jaya_rama_krishnaiah_2016}
\bibfield{author}{\bibinfo{person}{Hari~Krishna Kanagala} {and} \bibinfo{person}{V.V. Jaya Rama~Krishnaiah}.} \bibinfo{year}{2016}\natexlab{}.
\newblock \bibinfo{title}{A Comparative Study of K-Means, DBSCAN and OPTICS}.
\newblock , \bibinfo{numpages}{6}~pages.
\newblock
\urldef\tempurl%
\url{https://doi.org/10.1109/ICCCI.2016.7479923}
\showDOI{\tempurl}


\bibitem[Karen and Sandra(2017)]%
        {karen2017contextual}
\bibfield{author}{\bibinfo{person}{Holtzblatt Karen} {and} \bibinfo{person}{Jones Sandra}.} \bibinfo{year}{2017}\natexlab{}.
\newblock \showarticletitle{Contextual inquiry: A participatory technique for system design}.
\newblock In \bibinfo{booktitle}{\emph{Participatory design}}. \bibinfo{publisher}{CRC Press}, \bibinfo{pages}{177--210}.
\newblock


\bibitem[Kim et~al\mbox{.}(2022)]%
        {Kim2022Aural}
\bibfield{author}{\bibinfo{person}{Sungyoung Kim}, \bibinfo{person}{Doyuen Ko}, \bibinfo{person}{Miriam~A. Kolar}, {and} \bibinfo{person}{Xuan Lu}.} \bibinfo{year}{2022}\natexlab{}.
\newblock \showarticletitle{Aural heritage preservation and access: Methodological explorations from data collection to immersive multimodal virtual reality}.
\newblock \bibinfo{journal}{\emph{The Journal of the Acoustical Society of America}} (\bibinfo{year}{2022}).
\newblock
\urldef\tempurl%
\url{https://doi.org/10.1121/10.0016251}
\showDOI{\tempurl}


\bibitem[Kliuchko et~al\mbox{.}(2019)]%
        {Kliuchko2019Fractionating}
\bibfield{author}{\bibinfo{person}{Marina Kliuchko}, \bibinfo{person}{Elvira Brattico}, \bibinfo{person}{Benjamin Gold}, \bibinfo{person}{Mari Tervaniemi}, \bibinfo{person}{Brigitte Bogert}, \bibinfo{person}{Petri Toiviainen}, {and} \bibinfo{person}{Peter Vuust}.} \bibinfo{year}{2019}\natexlab{}.
\newblock \showarticletitle{Fractionating auditory priors: A neural dissociation between active and passive experience of musical sounds}.
\newblock \bibinfo{journal}{\emph{PLOS ONE}}  \bibinfo{volume}{14} (\bibinfo{date}{05} \bibinfo{year}{2019}), \bibinfo{pages}{e0216499}.
\newblock
\urldef\tempurl%
\url{https://doi.org/10.1371/journal.pone.0216499}
\showDOI{\tempurl}


\bibitem[Kong(2024)]%
        {Kong_VRGAME_Diabolo_ICH}
\bibfield{author}{\bibinfo{person}{Cuiting Kong}.} \bibinfo{year}{2024}\natexlab{}.
\newblock \showarticletitle{Digital Diabolo: A Virtual Reality Game for the Presentation of Intangible Cultural Heritage Through Participatory Design}. In \bibinfo{booktitle}{\emph{Proceedings of the Participatory Design Conference 2024: Situated Actions, Doctoral Colloquium, PDC Places, Communities - Volume 3}} (Sibu, Malaysia) \emph{(\bibinfo{series}{PDC '24})}. \bibinfo{publisher}{Association for Computing Machinery}, \bibinfo{address}{New York, NY, USA}, \bibinfo{pages}{19–23}.
\newblock
\showISBNx{9798400706554}
\urldef\tempurl%
\url{https://doi.org/10.1145/3661456.3666052}
\showDOI{\tempurl}


\bibitem[Kullback and Leibler(1951)]%
        {kullback1951information}
\bibfield{author}{\bibinfo{person}{Solomon Kullback} {and} \bibinfo{person}{Richard~A Leibler}.} \bibinfo{year}{1951}\natexlab{}.
\newblock \showarticletitle{On information and sufficiency}.
\newblock \bibinfo{journal}{\emph{The annals of mathematical statistics}} \bibinfo{volume}{22}, \bibinfo{number}{1} (\bibinfo{year}{1951}), \bibinfo{pages}{79--86}.
\newblock


\bibitem[Lab(2022)]%
        {amtlab2022genz}
\bibfield{author}{\bibinfo{person}{AMT Lab}.} \bibinfo{year}{2022}\natexlab{}.
\newblock \bibinfo{title}{The challenge to keep Gen Z interested in long-form, high-quality content}.
\newblock \bibinfo{howpublished}{Arts Management and Technology Laboratory (AMT Lab), Carnegie Mellon University}.
\newblock
\urldef\tempurl%
\url{https://amt-lab.org/blog/2022/10/the-challenge-to-keep-gen-z-interested-in-long-form-high-quality-content}
\showURL{%
\tempurl}


\bibitem[Labs(2024)]%
        {skybox2024}
\bibfield{author}{\bibinfo{person}{Blockade Labs}.} \bibinfo{year}{2024}\natexlab{}.
\newblock \bibinfo{title}{Skybox AI}.
\newblock \bibinfo{howpublished}{Blockade Labs, generating 360\textdegree panoramic images in glorious 8K resolution}.
\newblock
\urldef\tempurl%
\url{https://skybox.blockadelabs.com/}
\showURL{%
\tempurl}


\bibitem[LC(2023)]%
        {lc_together_2023}
\bibfield{author}{\bibinfo{person}{RAY LC}.} \bibinfo{year}{2023}\natexlab{}.
\newblock \showarticletitle{{TOGETHER} {ENOUGH}: {Collaborative} {Constructions} of {Adaptations} to {Climate} {Futures}}. In \bibinfo{booktitle}{\emph{Companion {Publication} of the 2023 {ACM} {Designing} {Interactive} {Systems} {Conference}}} \emph{(\bibinfo{series}{{DIS} '23 {Companion}})}. \bibinfo{publisher}{Association for Computing Machinery}, \bibinfo{address}{New York, NY, USA}, \bibinfo{pages}{55--59}.
\newblock
\showISBNx{978-1-4503-9898-5}
\urldef\tempurl%
\url{https://doi.org/10.1145/3563703.3596805}
\showDOI{\tempurl}


\bibitem[LC et~al\mbox{.}(2024)]%
        {lc_time_2024}
\bibfield{author}{\bibinfo{person}{RAY LC}, \bibinfo{person}{Sijia Liu}, \bibinfo{person}{Latisha~Besariani Hendra}, {and} \bibinfo{person}{Kexue Fu}.} \bibinfo{year}{2024}\natexlab{}.
\newblock \showarticletitle{{TIME} {ENOUGH}: {Generative} {AI} {Visions} of {Climate} {Change} as {Cave} {Paintings} of the {Future}}. In \bibinfo{booktitle}{\emph{Proceedings of the 16th {Conference} on {Creativity} \& {Cognition}}} \emph{(\bibinfo{series}{C\&amp;{C} '24})}. \bibinfo{publisher}{Association for Computing Machinery}, \bibinfo{address}{New York, NY, USA}, \bibinfo{pages}{608--613}.
\newblock
\showISBNx{9798400704857}
\urldef\tempurl%
\url{https://doi.org/10.1145/3635636.3672190}
\showDOI{\tempurl}


\bibitem[LC et~al\mbox{.}(2023a)]%
        {lc_active_2023}
\bibfield{author}{\bibinfo{person}{RAY LC}, \bibinfo{person}{Sijia Liu}, {and} \bibinfo{person}{Qiaosheng Lyu}.} \bibinfo{year}{2023}\natexlab{a}.
\newblock \showarticletitle{{IN}/{ACTive}: {A} {Distance}-{Technology}-{Mediated} {Stage} for {Performer}-{Audience} {Telepresence} and {Environmental} {Control}}. In \bibinfo{booktitle}{\emph{Proceedings of the 31st {ACM} {International} {Conference} on {Multimedia}}} \emph{(\bibinfo{series}{{MM} '23})}. \bibinfo{publisher}{Association for Computing Machinery}, \bibinfo{address}{New York, NY, USA}, \bibinfo{pages}{6989--6997}.
\newblock
\showISBNx{9798400701085}
\urldef\tempurl%
\url{https://doi.org/10.1145/3581783.3613791}
\showDOI{\tempurl}


\bibitem[LC et~al\mbox{.}(2023b)]%
        {lc_contradiction_2023}
\bibfield{author}{\bibinfo{person}{RAY LC}, \bibinfo{person}{Sihuang Man}, \bibinfo{person}{Xiying Bao}, \bibinfo{person}{Jinhan Wan}, \bibinfo{person}{Bo Wen}, {and} \bibinfo{person}{Zijing Song}.} \bibinfo{year}{2023}\natexlab{b}.
\newblock \showarticletitle{"{Contradiction} pushes me to improvise": {Performer} {Expressivity} and {Engagement} in {Distanced} {Movement} {Performance} {Paradigms}}.
\newblock \bibinfo{journal}{\emph{Proceedings of the ACM on Human-Computer Interaction}} \bibinfo{volume}{7}, \bibinfo{number}{CSCW2} (\bibinfo{date}{Oct.} \bibinfo{year}{2023}), \bibinfo{pages}{333:1--333:26}.
\newblock
\urldef\tempurl%
\url{https://doi.org/10.1145/3610182}
\showDOI{\tempurl}


\bibitem[LC and Mizuno(2021)]%
        {lc_designing_2021}
\bibfield{author}{\bibinfo{person}{RAY LC} {and} \bibinfo{person}{Daijiro Mizuno}.} \bibinfo{year}{2021}\natexlab{}.
\newblock \showarticletitle{Designing for {Narrative} {Influence}: {Speculative} {Storytelling} for {Social} {Good} in {Times} of {Public} {Health} and {Climate} {Crises}}.
\newblock In \bibinfo{booktitle}{\emph{Extended {Abstracts} of the 2021 {CHI} {Conference} on {Human} {Factors} in {Computing} {Systems}}}. Number~29. \bibinfo{publisher}{Association for Computing Machinery}, \bibinfo{address}{New York, NY, USA}, \bibinfo{pages}{1--13}.
\newblock
\showISBNx{978-1-4503-8095-9}
\urldef\tempurl%
\url{https://doi.org/10.1145/3411763.3450373}
\showURL{%
\tempurl}


\bibitem[LC et~al\mbox{.}(2022a)]%
        {lc_designing_2022}
\bibfield{author}{\bibinfo{person}{RAY LC}, \bibinfo{person}{Zijing Song}, \bibinfo{person}{Yating Sun}, {and} \bibinfo{person}{Cheng Yang}.} \bibinfo{year}{2022}\natexlab{a}.
\newblock \showarticletitle{Designing narratives and data visuals in comic form for social influence in climate action}.
\newblock \bibinfo{journal}{\emph{Frontiers in Psychology}}  \bibinfo{volume}{13} (\bibinfo{year}{2022}).
\newblock
\showISSN{1664-1078}
\urldef\tempurl%
\url{https://www.frontiersin.org/articles/10.3389/fpsyg.2022.893181}
\showURL{%
\tempurl}


\bibitem[LC et~al\mbox{.}(2022b)]%
        {lc_case_2022}
\bibfield{author}{\bibinfo{person}{RAY LC}, \bibinfo{person}{Anika Ullah}, {and} \bibinfo{person}{Fabeha Monir}.} \bibinfo{year}{2022}\natexlab{b}.
\newblock \showarticletitle{A {Case} for {Play}: {Immersive} {Storytelling} of {Rohingya} {Refugee} {Experience}}. In \bibinfo{booktitle}{\emph{International {Symposium} on {Electronic} {Art}}}. \bibinfo{address}{Barcelona, Spain}.
\newblock
\urldef\tempurl%
\url{https://isea2022.isea-international.org/event/full-paper-a-case-for-play-immersive-storytelling-of-rohingya-refugee-experience/}
\showURL{%
\tempurl}


\bibitem[Ling et~al\mbox{.}(2024)]%
        {ling_sketchar_2024}
\bibfield{author}{\bibinfo{person}{Long Ling}, \bibinfo{person}{Xinyi Chen}, \bibinfo{person}{Ruoyu Wen}, \bibinfo{person}{Toby Jia-Jun Li}, {and} \bibinfo{person}{RAY LC}.} \bibinfo{year}{2024}\natexlab{}.
\newblock \showarticletitle{Sketchar: {Supporting} {Character} {Design} and {Illustration} {Prototyping} {Using} {Generative} {AI}}.
\newblock \bibinfo{journal}{\emph{Proc. ACM Hum.-Comput. Interact.}} \bibinfo{volume}{8}, \bibinfo{number}{CHI PLAY} (\bibinfo{date}{Oct.} \bibinfo{year}{2024}), \bibinfo{pages}{337:1--337:28}.
\newblock
\urldef\tempurl%
\url{https://doi.org/10.1145/3677102}
\showDOI{\tempurl}


\bibitem[Liu et~al\mbox{.}(2023)]%
        {liu2023digital}
\bibfield{author}{\bibinfo{person}{Guanhong Liu}, \bibinfo{person}{Xianghua Ding}, \bibinfo{person}{Jinghe Cai}, \bibinfo{person}{Weiyun Wang}, \bibinfo{person}{Xinyue Wang}, \bibinfo{person}{Yuting Diao}, \bibinfo{person}{Jin Chen}, \bibinfo{person}{Tianyu Yu}, \bibinfo{person}{Haiqing Xu}, {and} \bibinfo{person}{Haipeng Mi}.} \bibinfo{year}{2023}\natexlab{}.
\newblock \showarticletitle{Digital making for inheritance and enlivening intangible cultural heritage: A case of hairy monkey handicrafts}. In \bibinfo{booktitle}{\emph{Proceedings of the 2023 CHI conference on human factors in computing systems}}. \bibinfo{pages}{1--17}.
\newblock


\bibitem[Liu et~al\mbox{.}(2025)]%
        {liu_salt_2025}
\bibfield{author}{\bibinfo{person}{Sijia Liu}, \bibinfo{person}{Xiaoke Zeng}, \bibinfo{person}{Fengyihan Wu}, \bibinfo{person}{Shu Ye}, \bibinfo{person}{Bowen Liu}, \bibinfo{person}{Sydney Cheung}, \bibinfo{person}{Richard~William Allen}, {and} \bibinfo{person}{RAY LC}.} \bibinfo{year}{2025}\natexlab{}.
\newblock \showarticletitle{"{Salt} is the {Soul} of {Hakka} {Baked} {Chicken}": {Reimagining} {Traditional} {Chinese} {Culinary} {ICH} for {Modern} {Contexts} {Without} {Losing} {Tradition}}. In \bibinfo{booktitle}{\emph{Creativity and {Cognition}}} \emph{(\bibinfo{series}{C\&{C} â€™25})}. \bibinfo{publisher}{Association for Computing Machinery}, \bibinfo{address}{New York, NY, USA}, \bibinfo{pages}{11}.
\newblock
\urldef\tempurl%
\url{https://doi.org/10.1145/3698061.3726917}
\showDOI{\tempurl}


\bibitem[Liu et~al\mbox{.}(2022)]%
        {Liu_Hua'er}
\bibfield{author}{\bibinfo{person}{Zixiao Liu}, \bibinfo{person}{Shuo Yan}, \bibinfo{person}{Yu Lu}, {and} \bibinfo{person}{Yuetong Zhao}.} \bibinfo{year}{2022}\natexlab{}.
\newblock \showarticletitle{Generating Embodied Storytelling and Interactive Experience of China Intangible Cultural Heritage “Hua'er” in Virtual Reality}. In \bibinfo{booktitle}{\emph{Extended Abstracts of the 2022 CHI Conference on Human Factors in Computing Systems}} (New Orleans, LA, USA) \emph{(\bibinfo{series}{CHI EA '22})}. \bibinfo{publisher}{Association for Computing Machinery}, \bibinfo{address}{New York, NY, USA}, Article \bibinfo{articleno}{439}, \bibinfo{numpages}{7}~pages.
\newblock
\showISBNx{9781450391566}
\urldef\tempurl%
\url{https://doi.org/10.1145/3491101.3519761}
\showDOI{\tempurl}


\bibitem[Lo(2012)]%
        {lo2012nonlinear}
\bibfield{author}{\bibinfo{person}{Charles Lo}.} \bibinfo{year}{2012}\natexlab{}.
\newblock \showarticletitle{Nonlinear dimensionality reduction for music feature extraction}.
\newblock \bibinfo{journal}{\emph{Tech. Rep. CSC2515}} (\bibinfo{year}{2012}).
\newblock


\bibitem[Lu et~al\mbox{.}(2011)]%
        {lu2011shadowstory}
\bibfield{author}{\bibinfo{person}{Fei Lu}, \bibinfo{person}{Feng Tian}, \bibinfo{person}{Yingying Jiang}, \bibinfo{person}{Xiang Cao}, \bibinfo{person}{Wencan Luo}, \bibinfo{person}{Guang Li}, \bibinfo{person}{Xiaolong Zhang}, \bibinfo{person}{Guozhong Dai}, {and} \bibinfo{person}{Hongan Wang}.} \bibinfo{year}{2011}\natexlab{}.
\newblock \showarticletitle{ShadowStory: creative and collaborative digital storytelling inspired by cultural heritage}. In \bibinfo{booktitle}{\emph{Proceedings of the SIGCHI Conference on Human Factors in Computing Systems}}. \bibinfo{pages}{1919--1928}.
\newblock


\bibitem[Lu et~al\mbox{.}(2019)]%
        {lu2019feel}
\bibfield{author}{\bibinfo{person}{Zhicong Lu}, \bibinfo{person}{Michelle Annett}, \bibinfo{person}{Mingming Fan}, {and} \bibinfo{person}{Daniel Wigdor}.} \bibinfo{year}{2019}\natexlab{}.
\newblock \showarticletitle{" I feel it is my responsibility to stream" Streaming and Engaging with Intangible Cultural Heritage through Livestreaming}. In \bibinfo{booktitle}{\emph{Proceedings of the 2019 CHI Conference on Human Factors in Computing Systems}}. \bibinfo{pages}{1--14}.
\newblock


\bibitem[Lungu(2022)]%
        {Lungu2022_QUALcoding}
\bibfield{author}{\bibinfo{person}{Maria Lungu}.} \bibinfo{year}{2022}\natexlab{}.
\newblock \showarticletitle{The Coding Manual for Qualitative Researchers}.
\newblock \bibinfo{journal}{\emph{American Journal of Qualitative Research}} \bibinfo{volume}{6}, \bibinfo{number}{1} (\bibinfo{date}{May} \bibinfo{year}{2022}), \bibinfo{pages}{232--237}.
\newblock
\urldef\tempurl%
\url{https://doi.org/10.29333/ajqr/12085}
\showDOI{\tempurl}


\bibitem[McFee et~al\mbox{.}(2015)]%
        {mcfee2015librosa}
\bibfield{author}{\bibinfo{person}{Brian McFee}, \bibinfo{person}{Colin Raffel}, \bibinfo{person}{Dawen Liang}, \bibinfo{person}{Daniel~PW Ellis}, \bibinfo{person}{Matt McVicar}, \bibinfo{person}{Eric Battenberg}, {and} \bibinfo{person}{Oriol Nieto}.} \bibinfo{year}{2015}\natexlab{}.
\newblock \showarticletitle{librosa: Audio and music signal analysis in python.}. In \bibinfo{booktitle}{\emph{SciPy}}. \bibinfo{pages}{18--24}.
\newblock


\bibitem[Mogensen({[n.\,d.]})]%
        {mogensenPROVOTYPINGAPPROACHSYSTEMS1992}
\bibfield{author}{\bibinfo{person}{Preben Mogensen}.} \bibinfo{year}{[n.\,d.]}\natexlab{}.
\newblock \showarticletitle{{{TOWARDS A PROVOTYPING APPROACH IN SYSTEMS DEVELOPMENT}}}.
\newblock  \bibinfo{volume}{4}, \bibinfo{number}{1} (\bibinfo{year}{[n.\,d.]}).
\newblock
\showISSN{1901-0990}


\bibitem[Mulauzi et~al\mbox{.}(2021)]%
        {Mulauzi2021Preservation}
\bibfield{author}{\bibinfo{person}{Felesia Mulauzi}, \bibinfo{person}{Phiri Bwalya}, \bibinfo{person}{Chishimba Soko}, \bibinfo{person}{Vincent Njobvu}, \bibinfo{person}{Jane Katema}, {and} \bibinfo{person}{Felix Silungwe}.} \bibinfo{year}{2021}\natexlab{}.
\newblock \showarticletitle{Preservation of audio-visual archives in Zambia}.
\newblock \bibinfo{journal}{\emph{ESARBICA Journal: Journal of the Eastern and Southern Africa Regional Branch of the International Council on Archives}} (\bibinfo{year}{2021}).
\newblock
\urldef\tempurl%
\url{https://doi.org/10.4314/esarjo.v40i.4}
\showDOI{\tempurl}


\bibitem[Muntean et~al\mbox{.}(2017)]%
        {muntean2017designing}
\bibfield{author}{\bibinfo{person}{Reese Muntean}, \bibinfo{person}{Alissa~N Antle}, \bibinfo{person}{Brendan Matkin}, \bibinfo{person}{Kate Hennessy}, \bibinfo{person}{Susan Rowley}, {and} \bibinfo{person}{Jordan Wilson}.} \bibinfo{year}{2017}\natexlab{}.
\newblock \showarticletitle{Designing cultural values into interaction}. In \bibinfo{booktitle}{\emph{Proceedings of the 2017 CHI Conference on Human Factors in Computing Systems}}. \bibinfo{pages}{6062--6074}.
\newblock


\bibitem[Nwabueze(2013)]%
        {Nwabueze2013}
\bibfield{author}{\bibinfo{person}{Caroline~Joelle Nwabueze}.} \bibinfo{year}{2013}\natexlab{}.
\newblock \showarticletitle{The role of intellectual property in safeguarding intangible cultural heritage in museums}.
\newblock \bibinfo{journal}{\emph{International journal of intangible heritage}}  \bibinfo{volume}{8} (\bibinfo{year}{2013}), \bibinfo{pages}{181--190}.
\newblock


\bibitem[Othman et~al\mbox{.}(2021)]%
        {Othman2021Overview}
\bibfield{author}{\bibinfo{person}{Razifah Othman}, \bibinfo{person}{Masitah Ahmad}, \bibinfo{person}{Othman Ibrahim}, \bibinfo{person}{Haziah Sa'ari}, \bibinfo{person}{Siti~Nuur–Ila Mat~Kamal}, {and} \bibinfo{person}{Aflah~Isa Darami}.} \bibinfo{year}{2021}\natexlab{}.
\newblock \showarticletitle{Overview of UX-UI Via Virtual Reality Project in Preserving the Intangible Cultural Heritage of Negeri Sembilan, Malaysia}.
\newblock  (\bibinfo{year}{2021}), \bibinfo{pages}{180--185}.
\newblock
\urldef\tempurl%
\url{https://doi.org/10.1109/ICSPC53359.2021.9689107}
\showDOI{\tempurl}


\bibitem[P{\'a}l and V{\'a}rkonyi(2020)]%
        {pal2020comparison}
\bibfield{author}{\bibinfo{person}{Tam{\'a}s P{\'a}l} {and} \bibinfo{person}{D{\'a}niel~T V{\'a}rkonyi}.} \bibinfo{year}{2020}\natexlab{}.
\newblock \showarticletitle{Comparison of Dimensionality Reduction Techniques on Audio Signals.}. In \bibinfo{booktitle}{\emph{ITAT}}. \bibinfo{pages}{161--168}.
\newblock


\bibitem[Qin and Karimi(2024)]%
        {Qin2023EXPRESS:}
\bibfield{author}{\bibinfo{person}{Yue Qin} {and} \bibinfo{person}{Hassan~A Karimi}.} \bibinfo{year}{2024}\natexlab{}.
\newblock \showarticletitle{Active and passive exploration for spatial knowledge acquisition: A meta-analysis}.
\newblock \bibinfo{journal}{\emph{Quarterly Journal of Experimental Psychology}} \bibinfo{volume}{77}, \bibinfo{number}{5} (\bibinfo{year}{2024}), \bibinfo{pages}{964--982}.
\newblock


\bibitem[Rao et~al\mbox{.}(2024)]%
        {rao2024formationcreator}
\bibfield{author}{\bibinfo{person}{Junkai Rao}, \bibinfo{person}{Feng Zhou}, \bibinfo{person}{Ju Dai}, \bibinfo{person}{Chi Li}, {and} \bibinfo{person}{Yong Hu}.} \bibinfo{year}{2024}\natexlab{}.
\newblock \showarticletitle{FormationCreator: Designing A VR Dance Formation System for Intangible Cultural Heritage Dance}. In \bibinfo{booktitle}{\emph{Extended Abstracts of the CHI Conference on Human Factors in Computing Systems}}. \bibinfo{pages}{1--7}.
\newblock


\bibitem[Remijn and Kojima(2010)]%
        {Remijn2010Active}
\bibfield{author}{\bibinfo{person}{Gerard~B Remijn} {and} \bibinfo{person}{Haruyuki Kojima}.} \bibinfo{year}{2010}\natexlab{}.
\newblock \showarticletitle{Active versus passive listening to auditory streaming stimuli: a near-infrared spectroscopy study}.
\newblock \bibinfo{journal}{\emph{Journal of biomedical optics}} \bibinfo{volume}{15}, \bibinfo{number}{3} (\bibinfo{year}{2010}), \bibinfo{pages}{037006--037006}.
\newblock


\bibitem[Schaeffer(2017)]%
        {Schaeffer_4listeningModes}
\bibfield{author}{\bibinfo{person}{Pierre Schaeffer}.} \bibinfo{year}{2017}\natexlab{}.
\newblock \bibinfo{booktitle}{\emph{6. The Four Listening Modes}}.
\newblock \bibinfo{publisher}{University of California Press}, \bibinfo{address}{Berkeley}, \bibinfo{pages}{80--93}.
\newblock
\showISBNx{9780520967465}
\urldef\tempurl%
\url{https://doi.org/doi:10.1525/9780520967465-012}
\showDOI{\tempurl}


\bibitem[Singh and Ghanges(2016)]%
        {singh2016speaker}
\bibfield{author}{\bibinfo{person}{Alka Singh} {and} \bibinfo{person}{Sureka Ghanges}.} \bibinfo{year}{2016}\natexlab{}.
\newblock \showarticletitle{SPEAKER RECOGNITION USING MFCC AND DELTADELTA MFCC AND CLASSIFICATION USING ARTIFICIAL NEURAL NETWORK}.
\newblock \bibinfo{journal}{\emph{2016 8th International Journal of Advance Research in Science and Engineering}} (\bibinfo{year}{2016}), \bibinfo{pages}{8354}.
\newblock


\bibitem[Sun and Wang(2024)]%
        {SUN_AR_ICH}
\bibfield{author}{\bibinfo{person}{Dongyan Sun} {and} \bibinfo{person}{Chengping Wang}.} \bibinfo{year}{2024}\natexlab{}.
\newblock \showarticletitle{Application of AR Technology in Intangible Cultural Heritage and Cultural Tourism}. In \bibinfo{booktitle}{\emph{Proceedings of the 3rd International Conference on Electronic Information Technology and Smart Agriculture}} (Sanya, China) \emph{(\bibinfo{series}{ICEITSA '23})}. \bibinfo{publisher}{Association for Computing Machinery}, \bibinfo{address}{New York, NY, USA}, \bibinfo{pages}{247–252}.
\newblock
\showISBNx{9798400716775}
\urldef\tempurl%
\url{https://doi.org/10.1145/3641343.3641387}
\showDOI{\tempurl}


\bibitem[Tan et~al\mbox{.}(2020)]%
        {Tan_ICH_WebAR}
\bibfield{author}{\bibinfo{person}{Peng Tan}, \bibinfo{person}{Damian Hills}, \bibinfo{person}{Yi Ji}, {and} \bibinfo{person}{Kaiping Feng}.} \bibinfo{year}{2020}\natexlab{}.
\newblock \showarticletitle{Case Study: Creating Embodied Interaction with Learning Intangible Cultural Heritage through WebAR}. In \bibinfo{booktitle}{\emph{Extended Abstracts of the 2020 CHI Conference on Human Factors in Computing Systems}} (Honolulu, HI, USA) \emph{(\bibinfo{series}{CHI EA '20})}. \bibinfo{publisher}{Association for Computing Machinery}, \bibinfo{address}{New York, NY, USA}, \bibinfo{pages}{1–6}.
\newblock
\showISBNx{9781450368193}
\urldef\tempurl%
\url{https://doi.org/10.1145/3334480.3375199}
\showDOI{\tempurl}


\bibitem[Tsita et~al\mbox{.}(2023)]%
        {Tsita2023A}
\bibfield{author}{\bibinfo{person}{Christina Tsita}, \bibinfo{person}{Maya Satratzemi}, \bibinfo{person}{Alexandros Pedefoudas}, \bibinfo{person}{Charalabos Georgiadis}, \bibinfo{person}{Maria Zampeti}, \bibinfo{person}{Evi Papavergou}, \bibinfo{person}{Syrago Tsiara}, \bibinfo{person}{Eleni Sismanidou}, \bibinfo{person}{Petros Kyriakidis}, \bibinfo{person}{Dionysios Kehagias}, {and} \bibinfo{person}{Dimitrios Tzovaras}.} \bibinfo{year}{2023}\natexlab{}.
\newblock \showarticletitle{A Virtual Reality Museum to Reinforce the Interpretation of Contemporary Art and Increase the Educational Value of User Experience}.
\newblock \bibinfo{journal}{\emph{Heritage}}  \bibinfo{volume}{6} (\bibinfo{date}{05} \bibinfo{year}{2023}), \bibinfo{pages}{4134--4172}.
\newblock
\urldef\tempurl%
\url{https://doi.org/10.3390/heritage6050218}
\showDOI{\tempurl}


\bibitem[{UNESCO}(2003)]%
        {unesco2003basictexts}
\bibfield{author}{\bibinfo{person}{{UNESCO}}.} \bibinfo{year}{2003}\natexlab{}.
\newblock \bibinfo{booktitle}{\emph{Basic Texts of the 2003 Convention for the Safeguarding of the Intangible Cultural Heritage}}.
\newblock \bibinfo{publisher}{UNESCO}, \bibinfo{address}{Paris}.
\newblock
\urldef\tempurl%
\url{https://ich.unesco.org/en/convention}
\showURL{%
\tempurl}


\bibitem[{UNESCO}(2008)]%
        {unesco2008Kunqu}
\bibfield{author}{\bibinfo{person}{{UNESCO}}.} \bibinfo{year}{2008}\natexlab{}.
\newblock \bibinfo{title}{Kun Qu opera - Intangible Cultural Heritage of Humanity}.
\newblock \bibinfo{howpublished}{\url{https://ich.unesco.org/en/RL/kun-qu-opera-00004}}.
\newblock


\bibitem[{UNESCO}(2010)]%
        {unesco2010peking}
\bibfield{author}{\bibinfo{person}{{UNESCO}}.} \bibinfo{year}{2010}\natexlab{}.
\newblock \bibinfo{title}{Peking Opera - Intangible Cultural Heritage of Humanity}.
\newblock \bibinfo{howpublished}{\url{https://ich.unesco.org/en/RL/peking-opera-00418}}.
\newblock


\bibitem[{UNESCO}(2023)]%
        {unesco_meshrep}
\bibfield{author}{\bibinfo{person}{{UNESCO}}.} \bibinfo{year}{2023}\natexlab{}.
\newblock \bibinfo{title}{Meshrep}.
\newblock \bibinfo{howpublished}{\url{https://ich.unesco.org/en/USL/meshrep-00304}}.
\newblock
\newblock
\shownote{Accessed: 2024-08-29}.


\bibitem[{UNESCO}(nd)]%
        {unesco_dive_ICHheritage_general_threats}
\bibfield{author}{\bibinfo{person}{{UNESCO}}.} \bibinfo{year}{n.d.}\natexlab{}.
\newblock \bibinfo{title}{Dive into Intangible Cultural Heritage}.
\newblock \bibinfo{howpublished}{\url{https://ich.unesco.org/en/dive}}.
\newblock
\newblock
\shownote{Accessed: 2024-08-29}.


\bibitem[{UNESCO Intangible Cultural Heritage}(2023)]%
        {unesco2023photoexhibition}
\bibfield{author}{\bibinfo{person}{{UNESCO Intangible Cultural Heritage}}.} \bibinfo{year}{2023}\natexlab{}.
\newblock \bibinfo{title}{We Are Living Heritage: Photo Exhibition 2023}.
\newblock \bibinfo{howpublished}{Online Exhibition}.
\newblock
\urldef\tempurl%
\url{https://ich.unesco.org/en/we-are-living-heritage-photo-exhibition-2023-01331}
\showURL{%
\tempurl}


\bibitem[Van~der Maaten and Hinton(2008)]%
        {van2008visualizing}
\bibfield{author}{\bibinfo{person}{Laurens Van~der Maaten} {and} \bibinfo{person}{Geoffrey Hinton}.} \bibinfo{year}{2008}\natexlab{}.
\newblock \showarticletitle{Visualizing data using t-SNE.}
\newblock \bibinfo{journal}{\emph{Journal of machine learning research}} \bibinfo{volume}{9}, \bibinfo{number}{11} (\bibinfo{year}{2008}).
\newblock


\bibitem[Vasilev(2023)]%
        {Vasilev_2023}
\bibfield{author}{\bibinfo{person}{Artem Vasilev}.} \bibinfo{year}{2023}\natexlab{}.
\newblock \bibinfo{title}{audio-splitter 0.1.0}.
\newblock
\newblock
\urldef\tempurl%
\url{https://github.com/temavasilev/audio-splitter}
\showURL{%
\tempurl}


\bibitem[VR(2019)]%
        {googlevr_2019}
\bibfield{author}{\bibinfo{person}{Google VR}.} \bibinfo{year}{2019}\natexlab{}.
\newblock \bibinfo{title}{Degrees of Freedom}.
\newblock
\newblock
\urldef\tempurl%
\url{https://developers.google.com/vr/discover/degrees-of-freedom}
\showURL{%
\tempurl}


\bibitem[Wensveen and Matthews({[n.\,d.]})]%
        {wensveenPrototypesPrototypingDesign2014}
\bibfield{author}{\bibinfo{person}{Stephan Wensveen} {and} \bibinfo{person}{Ben Matthews}.} \bibinfo{year}{[n.\,d.]}\natexlab{}.
\newblock \showarticletitle{Prototypes and Prototyping in Design Research}.
\newblock In \bibinfo{booktitle}{\emph{The {{Routledge Companion}} to {{Design Research}}} (\bibinfo{edition}{1} ed.)}, \bibfield{editor}{\bibinfo{person}{Paul~A. Rodgers} {and} \bibinfo{person}{Joyce Yee}} (Eds.). \bibinfo{publisher}{Routledge}, \bibinfo{pages}{262--276}.
\newblock
\showISBNx{978-1-315-75846-6}
\urldef\tempurl%
\url{https://doi.org/10.4324/9781315758466-25}
\showDOI{\tempurl}


\bibitem[Withers(1980)]%
        {Withers1980Unbalanced}
\bibfield{author}{\bibinfo{person}{Glenn~A Withers}.} \bibinfo{year}{1980}\natexlab{}.
\newblock \showarticletitle{Unbalanced growth and the demand for performing arts: An econometric analysis}.
\newblock \bibinfo{journal}{\emph{Southern Economic Journal}} (\bibinfo{year}{1980}), \bibinfo{pages}{735--742}.
\newblock


\bibitem[Yang et~al\mbox{.}(2022)]%
        {yang_ai_2022}
\bibfield{author}{\bibinfo{person}{Daijin Yang}, \bibinfo{person}{Yanpeng Zhou}, \bibinfo{person}{Zhiyuan Zhang}, \bibinfo{person}{Toby Jia-Jun Li}, {and} \bibinfo{person}{RAY LC}.} \bibinfo{year}{2022}\natexlab{}.
\newblock \showarticletitle{{AI} as an {Active} {Writer}: {Interaction} strategies with generated text in human-{AI} collaborative fiction writing}. In \bibinfo{booktitle}{\emph{Joint {Proceedings} of the {IUI} 2022 {Workshops}: {APEx}-{UI}, {HAI}-{GEN}, {HEALTHI}, {HUMANIZE}, {TExSS}, {SOCIALIZE}}}. \bibinfo{publisher}{CEUR-WS Team}, \bibinfo{pages}{56--65}.
\newblock
\urldef\tempurl%
\url{https://scholars.cityu.edu.hk/en/publications/publication(d901f5a2-0600-422f-b588-db5a59871961).html}
\showURL{%
\tempurl}


\bibitem[Yao et~al\mbox{.}(2024)]%
        {yao2024shadowmaker}
\bibfield{author}{\bibinfo{person}{Zhihao Yao}, \bibinfo{person}{Shiqing Lyu}, \bibinfo{person}{Yao Lu}, \bibinfo{person}{Qirui Sun}, \bibinfo{person}{Hanxuan Li}, \bibinfo{person}{Xuezhu Wang}, \bibinfo{person}{Guanhong Liu}, {and} \bibinfo{person}{Haipeng Mi}.} \bibinfo{year}{2024}\natexlab{}.
\newblock \showarticletitle{ShadowMaker: Sketch-Based Creation Tool for Digital Shadow Puppetry}. In \bibinfo{booktitle}{\emph{Extended Abstracts of the CHI Conference on Human Factors in Computing Systems}}. \bibinfo{pages}{1--5}.
\newblock


\bibitem[YouTube(2012)]%
        {kunqu_1699peachblossom}
\bibfield{author}{\bibinfo{person}{YouTube}.} \bibinfo{year}{2012}\natexlab{}.
\newblock \bibinfo{title}{"1699 Peach Blossom"}.
\newblock \bibinfo{howpublished}{\url{https://www.youtube.com/watch?v=CUPkUkLT69Y&t=2087s}}.
\newblock
\newblock
\shownote{Uploaded on January 15, 2012, with 115,724 views as of Jan 16, 2025}.


\bibitem[YouTube(2014)]%
        {qin_xianglian_2014}
\bibfield{author}{\bibinfo{person}{YouTube}.} \bibinfo{year}{2014}\natexlab{}.
\newblock \bibinfo{title}{"Qin Xianglian" [English Subtitles]}.
\newblock \bibinfo{howpublished}{\url{https://www.youtube.com/watch?v=5NzBu5v-ISE}}.
\newblock
\newblock
\shownote{Uploaded on July 18, 2014, with 454,924 views as of August 29, 2024}.


\bibitem[Yu et~al\mbox{.}(2021)]%
        {Yu_CHI2021_GuQin}
\bibfield{author}{\bibinfo{person}{Minjing Yu}, \bibinfo{person}{Meng Zhang}, \bibinfo{person}{Chun Yu}, \bibinfo{person}{Xiaoguang Ma}, \bibinfo{person}{Xing-Dong Yang}, {and} \bibinfo{person}{Jiawan Zhang}.} \bibinfo{year}{2021}\natexlab{}.
\newblock \showarticletitle{We Can Do More to Save Guqin: Design and Evaluate Interactive Systems to Make Guqin More Accessible to the General Public}. In \bibinfo{booktitle}{\emph{Proceedings of the 2021 CHI Conference on Human Factors in Computing Systems}} (Yokohama, Japan) \emph{(\bibinfo{series}{CHI '21})}. \bibinfo{publisher}{Association for Computing Machinery}, \bibinfo{address}{New York, NY, USA}, Article \bibinfo{articleno}{294}, \bibinfo{numpages}{12}~pages.
\newblock
\showISBNx{9781450380966}
\urldef\tempurl%
\url{https://doi.org/10.1145/3411764.3445175}
\showDOI{\tempurl}


\bibitem[Zeng et~al\mbox{.}(2025)]%
        {zeng_ronaldos_2025}
\bibfield{author}{\bibinfo{person}{Yuhan Zeng}, \bibinfo{person}{Yingxuan Shi}, \bibinfo{person}{Xuehan Huang}, \bibinfo{person}{Fiona Nah}, {and} \bibinfo{person}{RAY LC}.} \bibinfo{year}{2025}\natexlab{}.
\newblock \showarticletitle{"{Ronaldo}'s a poser!": {How} the {Use} of {Generative} {AI} {Shapes} {Debates} in {Online} {Forums}}. In \bibinfo{booktitle}{\emph{Proceedings of the {CHI} {Conference} on {Human} {Factors} in {Computing} {Systems}}} \emph{(\bibinfo{series}{{CHI} '25})}. \bibinfo{publisher}{Association for Computing Machinery}, \bibinfo{address}{New York, NY, USA}, \bibinfo{pages}{18}.
\newblock
\urldef\tempurl%
\url{https://doi.org/10.1145/3706598.3713829}
\showDOI{\tempurl}


\bibitem[Zhang et~al\mbox{.}(2023b)]%
        {Zhang2023A}
\bibfield{author}{\bibinfo{person}{Lufang Zhang}, \bibinfo{person}{Yue Wang}, \bibinfo{person}{Zhichuan Tang}, \bibinfo{person}{Xia Liu}, {and} \bibinfo{person}{Moran Zhang}.} \bibinfo{year}{2023}\natexlab{b}.
\newblock \showarticletitle{A Virtual Experience System of Bamboo Weaving for Sustainable Research on Intangible Cultural Heritage Based on VR Technology}.
\newblock \bibinfo{journal}{\emph{Sustainability}} (\bibinfo{year}{2023}).
\newblock
\urldef\tempurl%
\url{https://doi.org/10.3390/su15043134}
\showDOI{\tempurl}


\bibitem[Zhang et~al\mbox{.}(2025)]%
        {zhang_can_2025}
\bibfield{author}{\bibinfo{person}{Qinshi Zhang}, \bibinfo{person}{Ruoyu Wen}, \bibinfo{person}{Latisha~Besariani Hendra}, \bibinfo{person}{Zijian Ding}, {and} \bibinfo{person}{RAY LC}.} \bibinfo{year}{2025}\natexlab{}.
\newblock \showarticletitle{Can {AI} {Prompt} {Humans}? {Multimodal} {Agents} {Prompt} {Players}' {Game} {Actions} and {Show} {Consequences} to {Raise} {Sustainability} {Awareness}}. In \bibinfo{booktitle}{\emph{Proceedings of the {CHI} {Conference} on {Human} {Factors} in {Computing} {Systems}}} \emph{(\bibinfo{series}{{CHI} '25})}. \bibinfo{publisher}{Association for Computing Machinery}, \bibinfo{address}{New York, NY, USA}, \bibinfo{pages}{29}.
\newblock
\urldef\tempurl%
\url{https://doi.org/10.1145/3706598.3713661}
\showDOI{\tempurl}


\bibitem[Zhang et~al\mbox{.}(2023a)]%
        {Zhangetal_VR_NVSHU_ICH}
\bibfield{author}{\bibinfo{person}{Xuanmiao Zhang}, \bibinfo{person}{Linqi Sun}, {and} \bibinfo{person}{Shuo Yan}.} \bibinfo{year}{2023}\natexlab{a}.
\newblock \showarticletitle{NVSHU: Virtual Reality Design and Narrative Popularization for Intangible Cultural Heritage Characters}. In \bibinfo{booktitle}{\emph{SIGGRAPH Asia 2023 XR}} (Sydney, NSW, Australia) \emph{(\bibinfo{series}{SA '23})}. \bibinfo{publisher}{Association for Computing Machinery}, \bibinfo{address}{New York, NY, USA}, Article \bibinfo{articleno}{22}, \bibinfo{numpages}{2}~pages.
\newblock
\showISBNx{9798400703164}
\urldef\tempurl%
\url{https://doi.org/10.1145/3610549.3614615}
\showDOI{\tempurl}


\bibitem[Zhou et~al\mbox{.}(2025)]%
        {zhou_retrochat_2025}
\bibfield{author}{\bibinfo{person}{Suifang Zhou}, \bibinfo{person}{Kexue Fu}, \bibinfo{person}{Huamin Yi}, {and} \bibinfo{person}{RAY LC}.} \bibinfo{year}{2025}\natexlab{}.
\newblock \showarticletitle{{RetroChat}: {Designing} for the {Preservation} of {Past} {Chinese} {Online} {Social} {Experiences}}. In \bibinfo{booktitle}{\emph{Creativity and {Cognition}}} \emph{(\bibinfo{series}{C\&{C} â€™25})}. \bibinfo{publisher}{Association for Computing Machinery}, \bibinfo{address}{New York, NY, USA}, \bibinfo{pages}{19}.
\newblock
\urldef\tempurl%
\url{https://doi.org/10.1145/3698061.3726920}
\showDOI{\tempurl}


\bibitem[Zhou et~al\mbox{.}(2024)]%
        {zhou_eternagram_2024}
\bibfield{author}{\bibinfo{person}{Suifang Zhou}, \bibinfo{person}{Latisha~Besariani Hendra}, \bibinfo{person}{Qinshi Zhang}, \bibinfo{person}{Jussi Holopainen}, {and} \bibinfo{person}{RAY LC}.} \bibinfo{year}{2024}\natexlab{}.
\newblock \showarticletitle{Eternagram: {Probing} {Player} {Attitudes} {Towards} {Climate} {Change} {Using} a {ChatGPT}-driven {Text}-based {Adventure}}. In \bibinfo{booktitle}{\emph{Proceedings of the {CHI} {Conference} on {Human} {Factors} in {Computing} {Systems}}} \emph{(\bibinfo{series}{{CHI} '24})}. \bibinfo{publisher}{Association for Computing Machinery}, \bibinfo{address}{New York, NY, USA}, \bibinfo{pages}{1--23}.
\newblock
\showISBNx{9798400703300}
\urldef\tempurl%
\url{https://doi.org/10.1145/3613904.3642850}
\showDOI{\tempurl}


\end{thebibliography}
\end{document}